\documentclass{sig-alternate}
\usepackage{kantlipsum}
\makeatletter
\def\@maketitle{\newpage
 \null
 \setbox\@acmtitlebox\vbox{%
\baselineskip 20pt
\vskip 2em                   
   \begin{center}
    {\ttlfnt \@title\par}       
    \vskip 1.5em                
{\subttlfnt \the\subtitletext\par}\vskip 1.25em
    {\baselineskip 16pt\aufnt   
     \lineskip .5em             
     \begin{tabular}[t]{c}\@author
     \end{tabular}\par}
    \vskip 1.5em               
   \end{center}}
 \dimen0=\ht\@acmtitlebox
\advance\dimen0 by -12pc\relax 
 \unvbox\@acmtitlebox
 \ifdim\dimen0<0.0pt\relax\vskip-\dimen0\fi}
\makeatother

\usepackage{amsfonts}
\usepackage{amsmath}
\usepackage{amssymb}
\usepackage{booktabs}
\usepackage{caption}
\usepackage{color}
\usepackage{enumitem}
\usepackage{graphicx}
\usepackage[utf8]{inputenc}
\usepackage{mathtools}
\usepackage{needspace}
\usepackage{pgfplots}
\usepackage{siunitx}
\usepackage{subcaption}
\usepackage{tikz}
\usepackage{tikz-qtree}
\usepackage{times}
\usepackage[hyphens]{url}
\usepackage{wrapfig}
\usepackage[activate={true,nocompatibility},final,tracking=true,kerning=true,spacing=true,factor=1100,stretch=10,shrink=10]{microtype}
\microtypecontext{spacing=nonfrench}

\usetikzlibrary{decorations.pathreplacing, patterns, shapes}

\definecolor{purple}{RGB}{255,187,255}
\definecolor{orange}{RGB}{255,218,185}
\definecolor{blue}{RGB}{170,249,255}
\definecolor{grey}{RGB}{160,160,160}

\newcommand{\para}[1]{\vspace{1mm} \noindent {\bf #1}}

\title{An Automated Social Graph De-anonymization Technique}
\numberofauthors{2}
\author{
    \alignauthor Kumar Sharad\\
    \affaddr{University of Cambridge, UK}\\
    \email{kumar.sharad@cl.cam.ac.uk}
    \alignauthor George Danezis\\
    \affaddr{University College London, UK}\\
    \email{g.danezis@ucl.ac.uk}
}

\usepackage{etoolbox}
\makeatletter
\patchcmd{\maketitle}{\@copyrightspace}{}{}{}
\makeatother

\begin{document}
    \sloppy
    \maketitle

    \begin{abstract}
        We present a generic and automated approach to re-identifying nodes in anonymized social networks which enables novel anonymization techniques to be quickly evaluated. It uses machine learning (decision forests) to matching pairs of nodes in disparate anonymized sub-graphs. The technique uncovers artefacts and invariants of any black-box anonymization scheme from a small set of examples. Despite a high degree of automation, classification succeeds with significant true positive rates even when small false positive rates are sought. Our evaluation uses publicly available real world datasets to study the performance of our approach against real-world anonymization strategies, namely the schemes used to protect datasets of The Data for Development (D4D) Challenge. We show that the technique is effective even when only small numbers of samples are used for training. Further, since it detects weaknesses in the black-box anonymization scheme it can re-identify nodes in one social network when trained on another.
    \end{abstract}


    \section{Introduction}\label{sec:Introduction}
        A number of rich datasets have recently been published for research purposes, often with only casual attempts to anonymize them. Research in de-anonymization has also seen an upswing~\cite{backstrom2007wherefore,conf/wpes/MulderDBP08,narayanan2011link,wondracek2010practical}, leading to high profile data releases being followed by high profile privacy breaches.

        These developments have forced organizations to make some effort to anonymize the released data. However, overly distorting data to achieve this contradicts the very purpose of a release, since it negatively impacts utility.

        Social network graphs in particular are high dimensional and feature rich data sets, and it is extremely hard to preserve their anonymity. Thus, any anonymization scheme has to be evaluated in detail, including those with a sound theoretical basis~\cite{Kifer:2011:NFL:1989323.1989345}. Techniques have been proposed to resist de-anonymization~\cite{Hay:2008:RSR:1453856.1453873,sala2011sharing,zhou2008preserving}, however, Dwork and Naor have shown~\cite{dwork2008difficulties} that preserving privacy of an individual whose data is released cannot be achieved in general.

        \para{Ad-hoc vs generic.} It has been conclusively demonstrated that merely removing identifiers in social network datasets is not sufficient to guarantee privacy. Despite these results, data practitioners, continue to propose anonymization strategies in the hope that they can resist de-anonymization ``in practice'', such as the ones used to protect datasets from The Data for Development (D4D) challenge. This has lead to a \emph{cat-and-mouse game}:
        Research thus far has focused on defeating new variants of anonymization techniques by manually devising ad-hoc de-anonymization techniques. Despite their simplicity, unraveling each anonymization technique manually requires considerable effort and time and each attack can be defeated by a small tweak to the anonymization strategy, often by destroying specific features on which the attack has been constructed. Tailoring attacks to specific scenarios~\cite{narayanan2008robust} highlights the problem of anonymization but the expense involved in evaluating each new scheme cannot be amortized.

        Better solutions are needed which attack entire classes of anonymization schemes rather than taking a piecewise approach. Such generic de-anonymization techniques will allow cheap and timely evaluation of novel anonymization schemes. In this paper, we demonstrate the efficacy of automated de-anonymization attacks on real-world anonymization schemes. They automatically uncover artefacts remaining after anonymization that allow for re-identification of nodes in social networks. The automated attacks can be used quickly and cheaply to demonstrate that a non negligible number of users would be at risk of de-anonymization for ``novel'' anonymization schemes. Specifically, we:
        \begin{itemize}[noitemsep,nolistsep]
            \item Formulate the problem of de-anonymization in social networks as a learning task. From a set of examples of known correspondences between nodes (\emph{training} data) we wish to learn a good de-anonymization model.

            \item Describe a non-parametric learning algorithm tailored to the de-anonymization learning problem in social graphs. The algorithm is based on random decision forests, with custom features that match social network nodes.

            \item Evaluate the learning algorithm on a real-world de-anonymization task from the D4D challenge, and compare it with an ad-hoc approach.

            \item Show that the algorithm and model learn sufficient information about the anonymization algorithm, rather than the specific dataset anonymized, to be useful when de-anonymizing social networks of a different nature than the ones used for training.

            \item We apply the automated learning algorithm, to a standard problem~\cite{narayanan2009anonymizing} of de-anonymizing nodes across social networks. It performs well, even when a very small number of examples are used to train it.
        \end{itemize}

        \para{Note on research ethics.}
        The experiments reported in this paper are performed on previously published social networks specifically shared for research by the Stanford Network Analysis Platform. We do not perform de-anonymization attacks on live networks, scraped data, or previously unlinkable datasets.


    \section{Motivation: The D4D Challenge}\label{sec:d4d}
        In July 2012 Orange unveiled the \emph{Data for Development} (D4D) challenge\footnote{\begin{scriptsize}\url{http://www.d4d.orange.com/}\end{scriptsize}} to promote research in behavioral science in the Ivory Coast using a number of mobile call datasets. The datasets are based on ``anonymized'' \emph{call detail records} extracted from Orange's customer base. The data was collected for 150 days, from December 1, 2011 until April 28, 2012, and contains 2.5 billion calls and SMS exchanges between around 5 million users, i.e.\ a quarter of the Ivory Coast population. Teams with access to the datasets had to sign appropriate non-disclosure and ethics agreements to protect privacy and supplement the anonymization procedures employed.

        Prior to release, the organizers asked us to evaluate the quality of the anonymization scheme used. In total four datasets were released for analysis~\cite{DBLP:journals/corr/abs-1210-0137}. In this work we concentrate on analyzing the anonymization procedures for \emph{Dataset 4}, representing an anonymized phone call graph.


        \subsection{Data Release and Anonymization}\label{sec:data release}
            The two variants of the anonymization process select a number of highly connected individuals (egos) and output a social sub-graph around them (egonets).

            \para{Scheme~1.} The initial dataset contains sub-graphs of about 8\,300 randomly selected individuals. Each sub-graph contains all communications amongst the egos and their contacts for up to 2 degrees of separation. The data also includes the volume of calls, duration of calls within a period and their directionality. The data spans a period of 150 days which is split into two week chunks. Individuals are assigned random identifiers which remain the same across time slots but are unique to each sub-graph. This is meant to obfuscate links between sub-graphs to prevent an adversary from reconstructing a larger communication graph by stitching sub-graphs together. We refer to this anonymization strategy as ``Scheme~1''.

            \para{Scheme~2.} Following our preliminary analysis the organizers independently modified the anonymization scheme prior to the final release: the number of egonets released was reduced from 8\,300 to 5\,000; all links between the 2-hop nodes (friend-of-friend to friend-of-friend of the ego) were deleted; the number of calls, duration of calls and the directionality of calls between two users was removed; calls with easy to re-identify features were removed, such as to or from public phones.
            These changes degrade the information available to adversaries but, as we demonstrate, they do not guarantee anonymity. We refer to the modified anonymization strategy as ``Scheme~2''.

            The toy example presented in Figures \ref{fig:scheme1} and \ref{fig:scheme2} illustrates the difference between the two schemes when anonymizing the same ego sub-graph.

            \begin{figure}
                \centering
                \begin{tikzpicture}[scale=0.35,auto=left]
                    \node (n1)[style={draw, circle, fill=purple, minimum size=6mm}] at (0,5) {ego};

                    \node (n2)[style={draw, circle, fill=orange, minimum size=5mm}] at (3,3) {};
                    \node [left] at (2.5,3) {1-hop};
                    \node (n3)[style={draw, circle, fill=orange, minimum size=5mm}] at (-3,3) {};
                    \node [below] at (-3,2.5) {1-hop};
                    \node (n4)[style={draw, circle, fill=orange, minimum size=5mm}] at (2.5,7.5) {};
                    \node [above] at (2.5,8) {1-hop};

                    \node (n5)[style={draw, circle, fill=blue, minimum size=2mm}] at (6,3){};
                    \node [below] at (6.3,2.7) {2-hop};
                    \node (n6)[style={draw, circle, fill=blue, minimum size=2mm}] at (2.5,0) {};
                    \node [right] at (2.8,0) {2-hop};
                    \node (n7)[style={draw, circle, fill=blue, minimum size=2mm}] at (-3.5,6.5) {};
                    \node [above] at (-3.7,6.8) {2-hop};
                    \node (n8)[style={draw, circle, fill=blue, minimum size=2mm}] at (6,6.5) {};
                    \node [above] at (6,6.7) {2-hop};

                    \foreach \from/\to in {n1/n2, n1/n3, n1/n4, n2/n4, n2/n5, n2/n6, n3/n6, n3/n7, n4/n7, n4/n8, n5/n6, n5/n8, n6/n7}
                    \draw (\from)[style={grey}] -- (\to);
                \end{tikzpicture}
                \caption{Scheme~1 preserves the complete 2-hop network}
                \label{fig:scheme1}
                \vspace{3mm}
                \begin{tikzpicture}[scale=0.35,auto=left]
                    \node (n1)[style={draw, circle, fill=purple, minimum size=6mm}] at (0,5) {ego};

                    \node (n2)[style={draw, circle, fill=orange, minimum size=5mm}] at (3,3) {};
                    \node [left] at (2.5,3) {1-hop};
                    \node (n3)[style={draw, circle, fill=orange, minimum size=5mm}] at (-3,3) {};
                    \node [below] at (-3,2.5) {1-hop};
                    \node (n4)[style={draw, circle, fill=orange, minimum size=5mm}] at (2.5,7.5) {};
                    \node [above] at (2.5,8) {1-hop};

                    \node (n5)[style={draw, circle, fill=blue, minimum size=2mm}] at (6,3){};
                    \node [below] at (6.3,2.7) {2-hop};
                    \node (n6)[style={draw, circle, fill=blue, minimum size=2mm}] at (2.5,0) {};
                    \node [right] at (2.8,0) {2-hop};
                    \node (n7)[style={draw, circle, fill=blue, minimum size=2mm}] at (-3.5,6.5) {};
                    \node [above] at (-3.7,6.8) {2-hop};
                    \node (n8)[style={draw, circle, fill=blue, minimum size=2mm}] at (6,6.5) {};
                    \node [above] at (6,6.7) {2-hop};

                    \foreach \from/\to in {n1/n2, n1/n3, n1/n4, n2/n4, n2/n5, n2/n6, n3/n6, n3/n7, n4/n7, n4/n8}
                    \draw (\from)[style={grey}] -- (\to);
                \end{tikzpicture}
                \caption{Scheme~2 removes edges between 2-hop nodes}
                \label{fig:scheme2}
            \end{figure}


        \subsection{Robustness of Anonymization}\label{sec:definition}
            The intent behind both anonymization schemes is to obscure the full social graph by only releasing unlinkable sub-graphs. Releasing the full graph would lead to de-anonymization risks similar to those presented against Netflix~\cite{narayanan2008robust} or AOL\footnote{\begin{scriptsize}\url{http://www.nytimes.com/2006/08/09/technology/09aol.html}\end{scriptsize}} datasets. The primary aim of our privacy evaluation is to determine the extent to which the same node in different anonymized sub-graphs may or may not be linkable. Special attention needs to be paid to the \emph{false positive rate}, namely when we label two nodes as identical when in fact they are not. Even a small false positive rate may in general introduce considerable noise in piecing together sub-graphs, since matching a node to a $n$-node sub-graph may only contain a single match but at least $n-1$ mismatches.

            \para{Success of de-anonymization.} The success of de-anonymization is measured by an individual's risk of re-identification~\cite{narayanan2008robust,narayanan2009anonymizing,wondracek2010practical}. To model this we measure the probability of re-linking an individual present in two social graphs purely based on topology. More formally, we measure the indistinguishability of an identical pair of individuals from a random pair. Members of the pair belong to the different social networks under evaluation.

            We note that we have concentrated on matching nodes between egonets as a measure of success of de-anonymization for D4D. This is the task we were required to perform when evaluating the actual D4D dataset on behalf of the competition organizers. However, this reduces to a lower bound on the probability of matching an anonymized egonet with a fuller social graph or a fragment of the social graph. One may simply take the full social graph and apply the anonymization procedure to generate egonets. Then our techniques can be applied to determine a match.

            Finally, our success rates need to be interpreted as lower bounds. We use features and heuristics that are not complete: it is possible that better ones may be found that improve the de-anonymization rate. This is particularly true of the heuristics used as part of our machine learning approach. Automatically trained classifiers are capable of out-performing humans in many tasks, but are limited when it comes to using higher level features. Thus our results, even when de-anonymization rates are low, can never be interpreted as a proof of security, only an illustration of vulnerability.

        \subsection{Ad-hoc De-anonymization}\label{sec:adhoc}
            We first present an ad-hoc attack on Scheme~1 that links nodes across egonets. We observe that we may encounter three distinct cases when linking nodes from two different egonets, that allow different types of analysis:
            \begin{description}[noitemsep,nolistsep]
                \item[Case~1: ``1-hop''.] Both nodes are at a distance of 1-hop from the ego, or are the ego in both sub-graphs;
                \item[Case~2: ``1,2-hop''.] Only one of the two nodes is 2-hops from the ego in one sub-graph;
                \item[Case~3: ``2-hop''.] Both nodes are at a distance of 2-hops from the ego in both sub-graphs.
            \end{description}
            Given a large egonet it is easy to detect which of the cases we are tackling by first identifying the ego. In a medium-sized sub-graph it is the single node from which all others are at most 2-hops away. If more than one node satisfies this relation we cannot detect the ego accurately. However, this does not affect the attack.

            The ad-hoc attack exploits the variability of degree distributions of nodes in social networks, which is known to approximately follow a power-law~\cite{Mislove-2007}. We observe that the entire 1-hop neighborhood of nodes that are friends of the ego, is preserved in Scheme~1 (since the full degree 2 graph is extracted). Therefore for Case~1 node pairs we can generate a \emph{signature} for nodes that is invariant under the anonymization procedure: the signature for each node consists of the sorted list of degrees of all the nodes in its \emph{1-hop} network. For example, if a 1-hop node has 3 friends with degree 19, 8, and 25 in its 1-hop network then its signature is \{3, 8, 19, 25\}. When a large signature of two 1-hop nodes, in different egonets, matches exactly we classify them as being the same node. We present an evaluation of the effectiveness of this attack in Section~\ref{sec:Evaluation}.

            \subsection{Limitations of Ad-hoc De-anonymization}\label{sec:adhoclimit}
                We tried to extend the attack to node pairs in Case~2 and Case~3 by using an approximate match on their common 1-hop nodes. Since considerable parts of the neighborhood of 2-hop nodes are deleted we cannot use an exact signature match on them. Perhaps unsurprisingly, we got mixed results for this approach and they were neither consistent nor robust. Extending the attack to Scheme~2 was even more restrictive, the exact signature match is not applicable for Case~1 due to the edges being deleted in relation to the ego, thus precluding any extension to other categories.

                The failure to generalize this approach is quite limiting: overlaps between egonets are likely to contain a 2-hop node rather than exclusively 1-hop nodes, since the number of 2-hop nodes is considerably larger. We do not discuss an ad-hoc extension of this attack any further, and simply consider it inapplicable for Scheme~1 Cases~2 and 3 and Scheme~2 completely.

                Overall, constructing such ad-hoc attacks is an expensive procedure. It requires \emph{manual} identification of some \emph{invariant property} between sub-graphs. It is also rare that such ad-hoc attacks make use of a combination of weak features -- since identifying them by hand would be a laborious process. A reliance on quasi-invariant features makes such ad-hoc attacks quite fragile. This is exemplified by the anonymization procedure of Scheme~2 that severely degrades features vital to the ad-hoc attack, namely the edges between the 2-hop nodes.

                This points to a more significant problem, relating to the \emph{economics of privacy research}: designing new variants of anonymization procedures is cheap; analyzing them requires manual labor to extract new complex invariants, even when the schemes are quite obviously leaking a lot of identifying information. Instead, we propose a general approach that applies to the analysis of anonymized social graphs. It uses machine learning to automatically learn invariants or informative relations between the same node in different egonets or sub-graphs.


    \section{Learning De-anonymization}\label{sec:ML}
        \subsection{De-anonymization: A Learning Problem}
            A large number of variants of anonymization algorithms can be devised at low cost through combinations of sampling, injecting random nodes, removing specific or random nodes, or picking specific sub-graph. Each variant would traditionally require careful manual analysis to devise an effective de-anonymization strategy. Thus the evaluation of such algorithms -- and the demonstration that they perform poorly -- is a labor intensive process. However we observe a key commonality of ``useful'' anonymization procedures: they need to \emph{preserve a rich set of generic features of the original graph, and its subgraphs, to provide an acceptable degree of utility to legitimate data users}. As a consequence a subset of those features may be automatically refined as identifiers to de-anonymize the nodes.

            We propose to replace the manual process of devising de-anonymization strategies by a generic learning algorithm. The learning algorithm takes an anonymization algorithm and automatically learns a model that allows the de-anonymization of a significant fraction of social network nodes. The learning algorithm does not require a full description of the anonymization scheme. Instead, we provide the learning algorithm with examples of network nodes that are meant to be unlinkable. The algorithm then automatically learns sets of features that constitute effective invariants that allow linking and de-anonymization.

            Two de-anonymization problems are considered in this paper within the generic learning framework:
            \begin{itemize}[noitemsep,nolistsep]
                \item {\bf Linkage between sub-nets:} Two nodes in two different anonymized sub-graphs are presented to the algorithm, which must decide whether they represent the same subject or not. This is the problem associated with linking ego-nets in the D4D challenge. This task forms the core of our evaluation.

                \item {\bf Linkage between raw and anonymized graphs:} A vertex in a raw graph and a vertex in an anonymized graph are presented, and the algorithm has to decide whether they represent the same user. This more traditional task is discussed in Section~\ref{sec:traditional}.
            \end{itemize}
            The models learned for the tasks above are in both cases presented with two target user nodes in two different sub-graphs. Ultimately, their task is to classify the two target graph nodes as representing the same user or not.

            \para{The role of training examples.} Our learning algorithm makes use of examples of anonymized graphs, to learn invariants that can be used to link nodes. These can be generated at will by applying a black-box anonymization algorithm. It is best if the training examples come from the same distribution as the target anonymized networks. For example, the D4D challenge publishes mobile phone call graphs, so using some mobile phone call graph to generate training examples is best. Finding \emph{some} example call records, and processing them to produce examples of anonymized sub-graphs is easy for a serious adversary. We note that the training examples do not have to contain any nodes that will be in the actual target anonymized graphs, unlike previous approaches~\cite{narayanan2009anonymizing} that required known \emph{seeds} to bridge anonymized and raw graphs. Thus, for example, one may use an available call sub-graph from one operator to train an attack on anonymized call graph from a competing operator. In Section~\ref{sec:samedistribution} we provide an extensive analysis in this setting, with very few training examples.

            However, we note that training examples uncover artefacts of the anonymization technique that enable attacks, not merely quirks of the specific training data. This is illustrated (Section~\ref{sec:diffdistribution}) by the fact that training examples from graphs of a very different nature, leads to de-anonymization success in other graphs.

        \subsection{Decision Trees and Random Forests}
            Our generic model for de-anonymization is represented as a collection (forest) of decision trees on graph features, which are learned using a random decision forest training algorithm. A detailed treatment of decision forests and their applications can be found in Criminisi et al.~\cite{Criminisi2011TR}. We present a brief overview of those techniques.

            A random decision forest is an ensemble of randomly trained decision trees~\cite{breiman1984classification, ho1995random}. The predictions of each tree are collated together resulting in a performance improvement.

            \para{Decision trees.} A decision tree is comprised of nodes and edges arranged in a hierarchical structure. We use binary trees in our work because using $n$-ary trees ($n>2$) does not provide significant accuracy benefits~\cite{Criminisi2011TR}. The branch nodes of the tree are called \emph{split nodes} and the terminal nodes are the \emph{leaf nodes}. A decision tree uses a weak classifier at each split node to route an item to the ``left'' or ``right'' child node based on its features. The item is routed down the tree, ultimately reaching a leaf node where one class, or a distribution over classes, is assigned. The features on which the split node predicates are defined is problem specific -- we propose features to represent an anonymized graph in Section~\ref{sec:features}.

            \para{Training and testing.} The tree training phase involves injecting labeled data into the tree to optimize the split parameters at the internal nodes and determining the leaf predictors. Each internal node of the tree is associated with a binary classifier also known as \emph{weak learner}, whose output decides the data split for a given parameter. Here $\mathcal{S}^L$ denotes the set of points passed to the left child of the split node, $\mathcal{S}^R$ denotes the set of points passed to the right child. We denote $\mathcal{S}$ as the set $\mathcal{S}^L \cup \mathcal{S}^R$ of all items used to train a split
            node.
            To train each split node we select the feature that maximizes the information gain objective function:
            \begin{align}\label{eq:infogain}
                I = H(\mathcal{S}) - \sum_{i \in \{L,R\}} \frac{\mid \mathcal{S}^i \mid}{\mid \mathcal{S} \mid} H(\mathcal{S}^i)
            \end{align}
            where $H(\mathcal{S})$ is the Shannon entropy of the labels in set $\mathcal{S}$ with elements belonging to the set of classes $\mathcal{C}$. Maximizing the information gain at every split node decreases entropy of data and increases confidence in tree prediction. Each leaf node stores the empirical distribution of the classes associated with the subset of training data reaching that leaf node. The predictor is defined as the probability of a data point belonging to a particular class based on the distribution. We only use binary classification, hence we have just two classes.

             \para{The decision forest model.}
            The trees of a random forest are trained independently in parallel. A data point is classified by pushing it through all the trees, branching left or right within each tree according to the item features, until it reaches the leaves of all trees (this process can also be parallelized). The prediction of each tree at the corresponding leaf node is averaged to generate an aggregate forest prediction, defined as
            \begin{align}\label{eq:forest prediction}
                p(c|v) = \frac{1}{T} \sum\limits_{t=1}^T p_t(c|v)
            \end{align}
            where $p(c|v)$ is the empirical probability (as computed by the random forest) that given a feature vector $v$, it belongs to a data point of class $c$, $T$ is the number of trees and $p_t(c|v)$ is the prediction of an individual tree. We use $T=400$ trees in our experiments.


        \subsection{Specialized Social Graph Features}\label{sec:features}
            Random forests rely on finding features that when combined using many weak learners, classify the data accurately. Training ensures that the best features are retained, and their best conjunction will be selected per tree. Different trees allow disjunctions of features to contribute to classification.

            A balance must be struck between too generic or too specific features. If the features are too generic, in that they appear in most sub-graphs, then classification will be akin to guessing. On the other hand if they are too specific and rely on intricate graph structures, then they may be seriously damaged during anonymization. In all cases efficient feature extraction is important to allow for fast training and evaluation.

            We use the degrees (number of friends) of the ``friends" of a node as features. Most anonymization strategies strive to preserve some utility of the data, and damaging such a fundamental property of graph elements \emph{unpredictably} has an averse affect on utility. Hence, the very fact node degrees are \emph{predictably} perturbed also allows us to mount an attack: we expect the distribution of degrees in the neighborhood of sub-graph nodes to be related between different ego-nets. Furthermore, neighborhood degree distribution can be efficiently computed.

            For each node in the social social network we define a feature vector $v = (c_0, c_1, \dots, c_{n-1}) \in \mathbb{Z}$ of length $n$ made up of components which are bins of size $b$. Each component represents the number of neighbors that have a degree in a particular range. The $i$th component is the count $c_i$ of the number of neighbors with degree such that $i \cdot b <$ degree $\le (i+1) \cdot b$, where $i \in \{0,n-1\}$. If the degree exceeds the maximum possible range of $n \cdot b$ then it is included in the last bin.

\begin{figure}[h!]
                \centering
                \resizebox{0.95\linewidth}{!}{
                \begin{tikzpicture}
                    \foreach \c/\i [count=\n] in {blue!20/$c_0 = 8$, blue!20/$c_1 = 4$, blue!20/$c_2 = 0$}
                        \node[draw,fill=\c,minimum height=0.8cm,minimum width = 1.5cm,xshift=\n*1.5cm](N\n){\i} ;

                    \draw[|<->|] (0.75,-0.6) -- (2.26,-0.6) node[pos=0.5,below]{\small{\texttt{size} b = $15$}};
                    \draw[|<->|] (0.75,0.6) -- (13.06,0.6) node[pos=0.5,above]{n = $70$ \texttt{bins}};

                    \node [tape,draw,minimum size=0.8cm,tape bend top=none,tape bend height=0.4cm,rotate=90] at (5.45,0)(t){};
                    \node [tape,draw,minimum size=0.8cm,tape bend top=none,tape bend height=0.4cm,rotate=270] at (7.45,0)(t){};

                    \node at (6.45,0.4) {\dots};
                    \node at (6.45,-0.4) {\dots};

                    \foreach \c/\i [count=\m] in  {blue!20/$c_{4} = 3$}
                        \node[draw,fill=\c,minimum height=0.8cm,minimum width = 1.5cm,xshift=6.9cm+\m*1.5cm](N\m){\i};

                    \node [tape,draw,minimum size=0.8cm,tape bend top=none,tape bend height=0.4cm,rotate=90] at (9.35,0)(t){};
                    \node [tape,draw,minimum size=0.8cm,tape bend top=none,tape bend height=0.4cm,rotate=270] at (11.35,0)(t){};

                    \node at (10.45,0.4) {\dots};
                    \node at (10.45,-0.4) {\dots};

                    \foreach \c/\i [count=\o] in  {blue!20/$c_{69} = 2$}
                        \node[draw,fill=\c,minimum height=0.8cm,minimum width = 1.5cm,xshift=10.8cm+\m*1.5cm](N\m){\i};
                \end{tikzpicture}}
                \caption{Example node feature vector}
                \label{fig:sample feature vector}
            \end{figure}
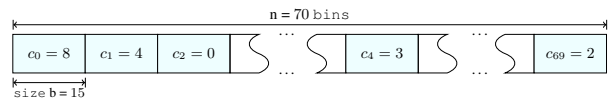

            Binning of neighborhood degrees is performed for efficiency and performance. Since the exact degree of the nodes across different anonymized sub-graphs may vary, the exact degree is irrelevant, and binning increases matching robustness.
            In our experiments we use $n=70$ and $b=15$. Figure \ref{fig:sample feature vector} illustrates the feature vector $v = (8,4,0,0,3,0,\dots,0,2)$ of a node with neighbors of degrees -- $\{1,1,3,3,5,6,7,13,16,20,21,30,65,69,72,1030,1100\}$.

            Apart from the depicted values all bins contain $0$'s.


        \subsection{Training and Classification of Node Pairs} \label{sec:training and classification}
            The basic data point to be classified consists of a pair of node features, each in a different egonet sub-graph. Let us denote the feature vector belonging to a node $n_p$ to be $v_p$. Each node pair $(n_p, n_q)$ is represented as a pair of vectors $(v_p, v_q)$. The node pair can be labeled as either ``identical'', if the nodes $n_p$ and $n_q$ are the same nodes in different egonets; or ``non-identical'' if $n_p$ and $n_q$ are in fact different nodes. The objective of the classification task is to infer the label of the pair $(n_p, n_q)$ on the basis of the feature vectors $v_p$ and $v_q$.

            \para{Training \& Weak Learner.} We use both bagging~\cite{breiman2001random} and randomized node optimization~\cite{ho1998random} to train decision trees. The trees are trained by injecting a random sample of data points at the root node. The sample contains data points of each class in equal proportion.

            The weak learner at each split node is presented with pairs of feature vectors $(v_p, v_q)$ and needs to decide whether they are assigned to the left or the right child node. To decide the split our weak learner uses the Silhouette Coefficient between two sample features $x \in v_p$ and $y \in v_q$ defined as:
            \begin{align}\label{eq:delta}
                \delta(x,y) =
                \begin{dcases*}
                    0 & if $x=y=0$\\
                    \frac{\mid x-y \mid}{\max(x,y)} & otherwise
                \end{dcases*}
            \end{align}
            where $x,y \in \mathbb{Z}$. Thus, for each feature pair $(v_p, v_q)$ we can calculate all $n^2$ component pairs $\delta(v_p[i],v_q[j])$, where $i,j \in \{0,n-1\}$. For a given component pair $(v_p[i],v_q[j])$ the tree computes $\delta(v_p[i],v_q[j])$ and passes the data point to the left or the right child depending upon threshold $\tau$. During training each split node is assigned a $\tau$ that splits the data for a given $(v_p[i], v_q[j])$ to maximize information gain. To choose the best $\tau$ for a given $(v_p[i], v_q[j])$ we cycle through $\tau \in [0,1]$ in steps of $0.05$. We inject randomness in the training of each split node by considering only a random 5\% of the total $n^2$ $(v_p[i],v_q[j])$ tuples of features.

            Once the values $(i,j, \tau)$ that best minimize entropy are determined for the split node, they are stored and do not change. The training procedure is repeated for its child nodes. We stop growing trees when the number of data points reaching a split node falls below 10\% of the total data points that were injected at the root node. This provides a good balance between trees that are too deep or too shallow, both of which provide poor results. Training ensures that the most informative features out of those available are learned at each split node. Randomizing the available set of features produces robust forests, that classify data according to a diverse set of mutually supporting features.

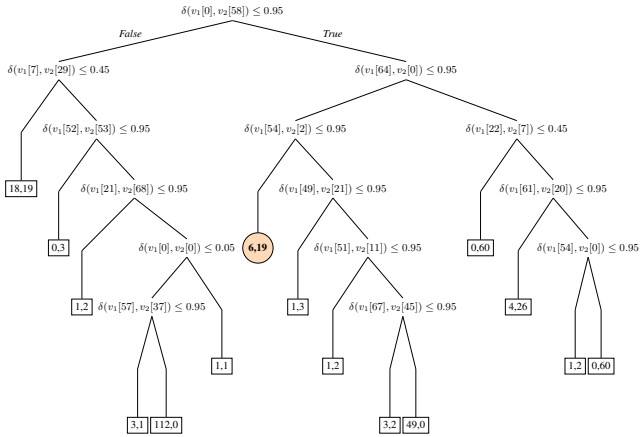
\begin{figure}
                \centering
                \resizebox{1.01\linewidth}{!}{
                \begin{tikzpicture}[level distance=50pt,sibling distance=2pt]
                \Tree [.$\delta(v_1[0],v_2[58])\le0.95$ \edge node[auto=right]{\emph{False}}; [.$\delta(v_1[7],v_2[29])\le0.45$ [\node[draw]{18,19}; ] [.$\delta(v_1[52],v_2[53])\le0.95$ [\node[draw]{0,3}; ] [.$\delta(v_1[21],v_2[68])\le0.95$ [\node[draw]{1,2}; ] [.$\delta(v_1[0],v_2[0])\le0.05$ [.$\delta(v_1[57],v_2[37])\le0.95$ [\node[draw]{3,1}; ] [\node[draw]{112,0}; ]] [\node[draw]{1,1}; ]]]]] \edge node[auto=left]{\emph{True}}; [.$\delta(v_1[64],v_2[0])\le0.95$ [.$\delta(v_1[54],v_2[2])\le0.95$ [\node[circle,draw,fill=orange]{{\bf 6,19}}; ] [.$\delta(v_1[49],v_2[21])\le0.95$ [\node[draw]{1,3}; ] [.$\delta(v_1[51],v_2[11])\le0.95$ [\node[draw]{1,2}; ] [.$\delta(v_1[67],v_2[45])\le0.95$ [\node[draw]{3,2}; ] [\node[draw]{49,0}; ]]]]] [.$\delta(v_1[22],v_2[7])\le0.45$ [\node[draw]{0,60}; ] [.$\delta(v_1[61],v_2[20])\le0.95$ [\node[draw]{4,26}; ] [.$\delta(v_1[54],v_2[0])\le0.95$ [\node[draw]{1,2}; ] [\node[draw]{0,60}; ]]]]]]
                \end{tikzpicture}}
                \caption{A randomly trained decision tree. The split nodes store weak learners $\delta(v_p[i],v_q[j]) \le \tau$, the highlighted leaf node has posterior of $(0.24,0.76)$}
                \label{fig:dectree}
            \end{figure}

            Figure \ref{fig:dectree} illustrates a sample decision tree: the split nodes contain the weak learner parameters, the left branches correspond to a \emph{False} result while the right branches correspond to \emph{True}. The leaf nodes store the count of (non-identical, identical) vector pairs reaching that leaf. We note that the sets of features selected are, perhaps surprisingly, not always from buckets close to each other. This is counter intuitive but yields great classification success, illustrating the difficulty of choosing such features manually.

            \para{Classification.} This is the simplest part of the algorithm. Once the decision forest has been trained, a previously unseen data point is classified by each tree till it reaches the leaf node and the values at all leaf nodes are recorded. At each leaf we calculate its probability of corresponding to an identical or non-identical node pair as their empirical distribution. For the highlighted leaf node in Figure \ref{fig:dectree} the empirical distribution is $(6, 19)$ and the probability of being classified as non-identical $= \frac{6}{6+19} = 0.24$ and probability of being classified as identical $= \frac{19}{6+19} = 0.76$. After traversing each tree in the forest all probabilities are averaged, as shown in equation \ref{eq:forest prediction}, to compute the final prediction.

            We discuss our choices of random forest algorithm parameters in Section~\ref{sec:param}.


    \section{Evaluation}\label{sec:Evaluation}
        We evaluate the classifier by training and testing on publicly available real world social networks. Our analysis is based on datasets obtained from the \emph{Stanford Network Analysis Platform}\footnote{\begin{scriptsize}\url{https://snap.stanford.edu/data/index.html}\end{scriptsize}}. We use two social networks to demonstrate our results: Epinions -- an online social trust network of a consumer review site and Pokec -- the most popular online social network in Slovakia. These different types of networks illustrate the applicability of our automated de-anonymization techniques across different types of social graphs.

        We use a very small partition of each of those networks (see Section \ref{sec:experimental}) to produce labeled training sets for the two anonymization schemes under consideration. Random forests are trained on the labeled training sets, and evaluated on separate test sets. Successful classification of test node pairs as ``identical'' or ``non-identical'' illustrates the success an adversary would have in using our techniques to attack the anonymization schemes.

        To fully characterize the difficulty of de-anonymizing each type of node pairs (1-hop, 1-to-2-hop or 2-hop, see Section~\ref{sec:adhoc}) we also consider each category separately. In those cases we set aside training and test data containing only pairs for that category. This does not provide the adversary with any advantage, since the category of a node and node pair can be inferred easily from the anonymized egonets. We also provide an aggregate analysis with node pairs from mixed categories. This is meant as a baseline representing the most generic attack, where the adversary is not making use of category information.

        \subsection{Experimental Setup}\label{sec:experimental}
            The D4D datasets motivating this study comprised 5\,000 egonets of nodes selected at random out of customer base of about 5 million. In line with this we generate 100 egonets for the Epinions dataset which has 75\,879 nodes and 1\,000 egonets for the Pokec dataset which has 1\,632\,803 nodes. We select those 2-hop egonets to have more than 400 nodes for Epinions and 800 for Pokec (to focus on ``important'' nodes as in previous work). We only study linkage of nodes with degree greater than 5; lower degree nodes have too little information to be meaningfully re-identified in bulk.

            Table \ref{tab:sample_epi_pok} (Appendix \ref{appendix:Sample Sizes}) lists the number of training links extracted from the training ego-nets for different node pair categories. We make no distinction between categories of non-identical node pairs. For each classifier we train 400 trees, each with 600 identical node pairs (of the appropriate category) and 600 non-identical node pairs (1200 in total).

            \noindent {\bf Figures of merit.} The output of a trained random forest classifier is a real value in $[0, 1]$, where a higher score denotes a higher probability of nodes being non-identical. Decisions based on different thresholds lead to a trade-off between \emph{True Positives} (TP) (the fraction of identical nodes that are labeled as identical), and \emph{False Positives} (FP) (the fraction of non-identical nodes that are erroneously classified as identical). For each classifier, each dataset and each scheme we present the true positive rate, for selected values of false positives rates (0.01\%, 0.1\%, 1\% 10\%, and 25\%). The Receiver Operating Characteristic (ROC) curves (Appendix \ref{appendix:ROC}) illustrate all trade-offs points between TP and FP. A summary of the quality of the classifier is provided by the ``area under the curve'' (AUC) of the ROC curve.

        \subsection{Results: Same Training Distribution}\label{sec:samedistribution}
            We first present the results of classification of identical versus non-identical nodes when the training data comes from the same distribution as the data to be classified.

            \noindent {\bf Epinions.} Table \ref{tab:adhoc}, summarizes the performance of the \emph{ad-hoc attack} in re-identifying node pairs of Case~1, Scheme~1. This hand crafted specialized attack leads to almost perfect results. On 900 identical node pairs it identified each one of them; on testing for 4907 non-identical node pairs the algorithm was correct 99.98\% of the time. This is almost perfect classification, but the technique cannot be generalized to other cases or schemes.

            \begin{table}[h]
                \caption{Success percentage for ad-hoc de-anonymization of Scheme~1 for Case~1 node pairs}
                \begin{center}
                    \parbox{\linewidth}{
                    \centering
                    \textbf{Success percentage}\\
                    \begin{tabular}{ l S[table-format=-4.2] S[table-format=-4.2]}
                        \toprule
                                         & {Identical}   & {Non-identical} \\
                        \midrule
                        Epinions         & 100           & 99.98  \\
                        Pokec            & 100           & 99.96  \\
                        \bottomrule
                    \end{tabular}
                    }
                \end{center}
                \label{tab:adhoc}
            \end{table}

            The automatic machine learning approach can be generalized to all cases of both schemes, as summarized in Table~\ref{tab:epi_fpvstp}. For the sake of direct comparison, for Scheme~1 using a fixed small false positive rate~(0.1\%) we can recover 90.39\% of Case~1 links; we also recover 46.82\% of Case~2 links, and 17.36\% of Case~3 links. The attack generalizes to Scheme~2 where we can recover 52.92\%, 25.95\% and 7.33\% of links in Cases 1, 2 and 3 respectively. Despite Scheme~2 leading to lower re-identification rates, for a fixed false positive rate, we observe they are never negligible, and a released dataset is likely to lead to the linkage of a significant number of individuals between ``unlikable'' sub-graphs. For even smaller false positives a significant number of Case~1 links are recoverable: 70.81\% for Scheme~1 and 35.08\% for Scheme~2. Case~2 linkage is also never negligible and higher than 11.37\% for both Schemes. If the adversary has any side information (such as seed nodes) that can be used to establish a good prior, a higher rate of false positives (1\%) can be tolerated with linkage rates of over 35.67\% even for the hardest Case~3 links.

            Figure \ref{fig:roc_epi_self} (Appendix \ref{appendix:ROC}) illustrates the ROC curves for Scheme~1 and Scheme~2 for all types of node pairs along with area under the ROC curve (AUC). We note that the AUC for Scheme~2 is consistently above 95\% for all link categories, higher than for Scheme~1 for Cases~2 and 3. This is due to the classifier performing significantly better for Scheme~2 when high false positive rates can be tolerated -- which may or may not be the case depending on adversary side information.

            Overall, the machine learning approach yields good results for Scheme~1, Cases~2 and 3 which the ad-hoc de-anonymization could not attack. More importantly, it yields good results for Scheme~2, on which the ad-hoc attack was not applicable. As expected, we get better results for the 1-hop category of node pairs, and for Scheme~1, which retain the most information from the original graph.

            \begin{table}[t!]
                \caption{Epinions (self-validation): \emph{False Positive} vs. \emph{True Positive} for both schemes}
                \begin{center}
                    \textbf{Scheme~1}\\
                    \begin{tabular}{ l S[table-format=-2.1] S[table-format=-2.1] S[table-format=-2.1] S[table-format=-2.1] S[table-format=-2.1]}
                        \toprule
                        False Positive   & {0.01\%}  & {0.1\%}  & {1\%}  & {10\%}  & {25\%}      \\
                        \midrule
                        1-hop               & 70.81 & 90.39 & 96.80 & 99.38 & 99.63   \\
                        1,2-hop             & 30.35 & 46.82 & 67.25 & 87.15 & 93.35   \\
                        2-hop               & 5.11  & 17.36 & 35.32 & 68.42 & 84.65   \\
                        Complete            & 4.41  & 17.76 & 35.67 & 68.08 & 83.79   \\
                        \bottomrule
                    \end{tabular}

                    \vspace{3mm}
                    \textbf{Scheme~2}\\
                    \begin{tabular}{ l S[table-format=-2.1] S[table-format=-2.1] S[table-format=-2.1] S[table-format=-2.1] S[table-format=-2.1]}
                        \toprule
                        False Positive   & {0.01\%}  & {0.1\%}   & {1\%}     & {10\%}     & {25\%}     \\
                        \midrule
                        1-hop               & 35.08  & 52.92 & 82.87 & 97.33 & 100.00  \\
                        1,2-hop             & 11.37  & 25.95 & 62.11 & 83.95 & 93.70   \\
                        2-hop               & 1.86   & 7.33  & 47.71 & 99.98 & 100.00  \\
                        Complete            & 0.34   & 5.89  & 36.33 & 94.99 & 98.62   \\
                        \bottomrule
                    \end{tabular}
                \end{center}
                \label{tab:epi_fpvstp}
            \end{table}

            \noindent {\bf Pokec.} The results on the Pokec network are summarized in Table~\ref{tab:pok_fpvstp}, and are similar to Epinions, which supports the general applicability of the attack. The ad-hoc attack against Scheme~1 gives almost perfect results. For Case~1 and 2 links the true positive linkage rate of the machine learning re-identification procedure is 42.92\% and 11.58\% for Scheme~1; for Scheme~2 they are 16.26\% and 6.41\% respectively (for a 0.1\% false positive rate). Interestingly, the linkage rate for Case~3, which was the lowest in the Epinions network, now outperforms Case~2 and is on par with Case~1 when higher false positive rates are tolerable. This is likely due to a lower degree of overlap in the neighborhood of non-identical nodes resulting from sampling links out of 1\,000 ego-nets rather than 100 as was the case for Epinions.

            Figure~\ref{fig:roc_pok_x} (Appendix \ref{appendix:ROC}) illustrates the ROC curves for the Pokec network and the AUC.

            \begin{table}[t!]
                \caption{Pokec (self-validation): \emph{False Positive} vs. \emph{True Positive} for both schemes}
                \begin{center}
                    \textbf{Scheme~1}\\
                    \begin{tabular}{ l S[table-format=-2.1] S[table-format=-2.1] S[table-format=-2.1] S[table-format=-2.1] S[table-format=-2.1]}
                        \toprule
                        False Positive   & {0.01\%}  & {0.1\%}  & {1\%}  & {10\%}  & {25\%}      \\
                        \midrule
                        1-hop               & 27.50  & 42.92 & 51.04 & 88.75 & 93.96  \\
                        1,2-hop             & 5.25   & 11.58 & 36.16 & 73.24 & 88.68  \\
                        2-hop               & 0.00   & 12.55 & 23.15 & 49.14 & 69.96  \\
                        Complete            & 0.01   & 10.44 & 20.48 & 47.60 & 68.36  \\
                        \bottomrule
                    \end{tabular}

                    \vspace{3mm}
                    \textbf{Scheme~2}\\
                    \begin{tabular}{ l S[table-format=-2.1] S[table-format=-2.1] S[table-format=-2.1] S[table-format=-2.1] S[table-format=-2.1]}
                        \toprule
                        False Positive   & {0.01\%}  & {0.1\%}   & {1\%}     & {10\%}     & {25\%}     \\
                        \midrule
                        1-hop               & 4.20   & 16.26 & 49.89 & 97.20 & 99.58  \\
                        1,2-hop             & 0.79   & 6.41  & 28.32 & 73.88 & 94.66  \\
                        2-hop               & 1.62   & 12.12 & 50.42 & 99.96 & 99.99  \\
                        Complete            & 0.68   & 6.12  & 21.14 & 64.12 & 86.10  \\
                        \bottomrule
                    \end{tabular}
                \end{center}
                \label{tab:pok_fpvstp}
            \end{table}

        \subsection{Results: Different Training Distribution} \label{sec:diffdistribution}
            In this section we evaluate the performance of the classifiers trained on totally different distributions. For this, we turn the classifier trained on the 100 Epinions ego-nets to classify test data from Pokec (Table~\ref{tab:pok_xval_fpvstp}), and vice versa, we use the classifier trained on 1\,000 Pokec ego-nets to classify the Epinions test data (Table \ref{tab:epi_xval_fpvstp}). The two networks are of a totally different nature: Epinions is a small web-of-trust on a consumer reviews site; Pokec is a much larger national social network.

            It is clear that a different training distribution has a detrimental effect on the quality of classification for very low false positive values. For a false positive rate of 0.1\% Case~1, 2 and 3 links in the Pokec network are linked at a rate of 27.29\%, 10.10\% and 4.18\% respectively for Scheme~1, and 5.40\%, 2.08\% and 13.57\% for Scheme~2 (Table~\ref{tab:pok_xval_fpvstp}). The results on the Epinions network see a similar fall (Table~\ref{tab:epi_xval_fpvstp}). Despite this, they never become small enough to guarantee that the likelihood of linking is negligible. For higher false positive rates (1\%) the linkage rate becomes significant again, particularly for Case~1 links, with success rates of 34.79\% (Scheme~1, Pokec), 12.29\% (Scheme~2, Pokec), 53.45\% (Scheme~1, Epinions) and 64.51\% (Scheme~2, Epinions). The linkage rate for the Pokec network under Scheme~2 is the largest for Case~3 links, namely 13.57\% (0.1\% FP), and a surprising 45.45\% (1\% FP).

            We conclude that while training on examples from a totally different distribution yields lower true positive rates for equivalent false positive rates, it does not lead to secure configurations of either anonymization Scheme~1 or 2. If data were to be released, a significant number of links, particularly Case~1 links, could be uncovered with a low error rate. Where the adversary can use side information to build better priors (using additional attributes or known seed nodes as suggested by previous research) a significant fraction of the links would be recovered.

            \begin{table}[h!]
                \caption{Epinions (x-validation): \emph{False Positive} vs. \emph{True Positive} for both schemes}
                \begin{center}
                    \textbf{Scheme~1}\\
                    \begin{tabular}{ l S[table-format=-2.1] S[table-format=-2.1] S[table-format=-2.1] S[table-format=-2.1] S[table-format=-2.1]}
                        \toprule
                        False Positive   & {0.01\%}  & {0.1\%}  & {1\%}  & {10\%}  & {25\%}      \\
                        \midrule
                        1-hop               & 17.86  & 43.60 & 53.45 & 74.63 & 85.34  \\
                        1,2-hop             & 2.71   & 6.11  & 19.79 & 57.65 & 78.54  \\
                        2-hop               & 0.13   & 0.68  & 5.99  & 35.99 & 64.20  \\
                        Complete            & 0.05   & 1.87  & 7.52  & 33.40 & 61.54  \\
                        \bottomrule
                    \end{tabular}

                    \vspace{3mm}
                    \textbf{Scheme~2}\\
                    \begin{tabular}{ l S[table-format=-2.1] S[table-format=-2.1] S[table-format=-2.1] S[table-format=-2.1] S[table-format=-2.1]}
                        \toprule
                        False Positive   & {0.01\%}  & {0.1\%}   & {1\%}     & {10\%}     & {25\%}     \\
                        \midrule
                        1-hop               & 1.44   & 6.56  & 64.51 & 93.64 & 99.69  \\
                        1,2-hop             & 0.57   & 3.27  & 23.80 & 82.46 & 91.93  \\
                        2-hop               & 0.03   & 1.17  & 9.72  & 99.69 & 99.99  \\
                        Complete            & 0.72   & 2.96  & 23.42 & 93.12 & 97.75  \\
                        \bottomrule
                    \end{tabular}
                \end{center}
                \label{tab:epi_xval_fpvstp}
            \end{table}

            \begin{table}[h!]
                \caption{Pokec (x-validation): \emph{False Positive} vs. \emph{True Positive} for both schemes}
                \begin{center}
                    \textbf{Scheme~1}\\
                    \begin{tabular}{ l S[table-format=-2.1] S[table-format=-2.1] S[table-format=-2.1] S[table-format=-2.1] S[table-format=-2.1]}
                        \toprule
                        False Positive   & {0.01\%}  & {0.1\%}  & {1\%}  & {10\%}  & {25\%}      \\
                        \midrule
                        1-hop               & 19.38  & 27.29 & 34.79 & 57.92 & 76.25  \\
                        1,2-hop             & 2.98   & 10.10 & 26.52 & 70.37 & 90.72  \\
                        2-hop               & 1.71   & 4.18  & 18.84 & 39.12 & 52.52  \\
                        Complete            & 1.89   & 4.05  & 16.83 & 36.81 & 50.76  \\
                        \bottomrule
                    \end{tabular}

                    \vspace{3mm}
                    \textbf{Scheme~2}\\
                    \begin{tabular}{ l S[table-format=-2.1] S[table-format=-2.1] S[table-format=-2.1] S[table-format=-2.1] S[table-format=-2.1]}
                        \toprule
                        False Positive   & {0.01\%}  & {0.1\%}   & {1\%}     & {10\%}     & {25\%}     \\
                        \midrule
                        1-hop               & 2.11   & 5.40  & 12.29 & 28.29 & 60.26  \\
                        1,2-hop             & 0.18   & 2.08  & 14.34 & 49.25 & 70.76  \\
                        2-hop               & 3.02   & 13.57 & 45.45 & 99.80 & 100.00  \\
                        Complete            & 1.00   & 5.61  & 19.22 & 56.90 & 72.76  \\
                        \bottomrule
                    \end{tabular}
                \end{center}
                \label{tab:pok_xval_fpvstp}
            \end{table}

        \subsection{Traditional De-anonymization Task}\label{sec:traditional}
            To assess the generality of the proposed algorithm we apply it to the traditional de-anonymization task, which requires mapping individuals between two different social networks. The adversary uses auxiliary information from a social network at their disposal to compromise privacy using the learned mappings from the released network. Narayanan and Shmatikov~\cite{narayanan2009anonymizing} present an attack based on topology to map node pairs across social networks that proceeds in two phases: First, the attacker manually maps a few ``seed'' nodes that are present in both the anonymized target graph and attacker's auxiliary graph; then a propagation phase begins which extends the seed mappings to new nodes based on topology. The new mapping is fed back to the algorithm, eventually resulting in re-identification of nodes across the two graphs.

            The performance of the algorithm is evaluated in~\cite{narayanan2009anonymizing} by synthetically generating target and auxiliary graphs from a real social network. To generate auxiliary and target graphs two subsets of nodes $V_1$ and $V_2$ are sampled from a real social network $G$ with $V$ nodes and $E$ edges, $G = (V,E)$. The overlap $\alpha_V$ between $V_1$ and $V_2$ is measured in terms of Jaccard Coefficient, defined for two sets $X$ and $Y$ as $JC(X,Y) = \frac{\left| X \cap Y \right|}{\left| X \cup Y \right|}$. To obtain a node overlap of $\alpha_V$ $V$ is partitioned randomly into three subsets $V_A$, $V_B$ and $V_C$ of size $\frac{1-\alpha_V}{2}\left| V \right|$, $\alpha_V\left|V \right|$ and $\frac{1-\alpha_V}{2}\left| V \right|$ respectively and set $V_1 = V_A \cup V_B$ and $V_2 = V_B \cup V_C$.

            Further noise is injected through edge perturbation -- deleting edges between nodes in $V_1$ and $V_2$ at random. This is done by making two copies of $E$ and independently deleting edges at random from each copy. The two copies are then projected on to $V_1$ and $V_2$ to obtain $E_1$ and $E_2$. Deleting a fraction $\beta$ of edges from each copy produces a fraction of $(1-\beta)^2$ common edges and a fraction of $1-\beta^2$ present in at least one copy. Thus the edge overlap is, $\alpha_E = \frac{(1-\beta)^2}{1-\beta^2}$, to obtain this edge overlap between subgraphs common to both graphs a fraction $\beta = \frac{1-\alpha_E}{1+\alpha_E}$ of edges must be deleted from each copy of $E$.

            Given the graphs $V_1$, $V_2$ we trained our classifier to distinguish a node pair as being identical or non-identical. A handful of seed mappings are used to train the learning algorithm. We note that unlike~\cite{narayanan2009anonymizing} which requires seeds to be part of a 4-clique and of high degree, our seed selection is unconditional. We tweaked the features slightly to take advantage of the global information: we add the 1-hop feature vector (see Section \ref{sec:features}) to the feature vector of the 2-hop neighborhood of the target node to produce a vector twice as long. Decision trees only consider component pairs from the corresponding neighborhood vector to decide the split, i.e.\ a feature of 1-hop neighborhood is never matched to a feature of 2-hop neighborhood. We use the Flickr social graph~\cite{Zafarani+Liu:2009} (also used in~\cite{narayanan2009anonymizing}) to generate $V_1$ and $V_2$ of size about 50\,000 nodes with an $\alpha_V = 0.25$. We train 400 trees with 5\,000 non-identical pairs and a varying number of identical pairs (seeds) -- 10, 50, 250, 1250, and carry out testing on 10\,000 pairs of each class.

            Table \ref{tab:flickr_pert} demonstrates that even with as few as 10 seeds we get significant true positive rate (16.74\%) for an appropriately low false positive rate (1\%). Increasing the number of seeds and edge overlap improves the results. The re-identification achieved by our classifier is not dependent on a critical mass of seed mappings, large scale re-identification in the algorithm presented by Narayanan and Shmatikov depends sharply on the number of seed mappings. Our classifier degrades gracefully as the number of seeds are decreased. The results are similar even when the classifier is trained with graphs generated using Epinions and then used to attack graphs generated using Flickr, this conclusively points towards learning de-anonymization independent of data being analyzed.

            We present a shoulder to shoulder comparison of our approach, hence the results presented use seed mappings between graphs generated using Flickr. (We note that the previous results~\cite{narayanan2009anonymizing} are reported as absolute success or error percentages on a global matching task, rather than a pair-wise matching task, and a specific dataset. They report $30.8\%$ of mappings being re-identified correctly, $12.1\%$ incorrectly, and $57\%$ not being identified. Sadly, we cannot compare the results directly, since we do not perform a global match.)

            \begin{table}[h!]
                \caption{Flickr (edge perturbation): \emph{False Positive} vs. \emph{True Positive}}
                \begin{center}
                    \textbf{Varying number of seeds ($\alpha_{E} = 0.25$)}\\
                    \begin{tabular}{ l S[table-format=-2.1] S[table-format=-2.1] S[table-format=-2.1] S[table-format=-2.1] S[table-format=-2.1]}
                        \toprule
                        False Positive   & {0.01\%}  & {0.1\%}  & {1\%}  & {10\%}  & {25\%}      \\
                        \midrule
                        10 seeds            & 0.14  & 3.23 & 16.74 & 52.39 & 78.39  \\
                        50 seeds            & 0.72  & 5.61 & 22.01 & 58.38 & 84.20  \\
                        250 seeds           & 3.68  & 7.19 & 22.32 & 58.61 & 85.14  \\
                        1250 seeds          & 0.86  & 5.97 & 23.52 & 60.26 & 86.43  \\
                        \bottomrule
                    \end{tabular}

                    \vspace{3mm}
                    \textbf{Varying edge overlap (50 seeds)}\\
                    \begin{tabular}{ l S[table-format=-2.1] S[table-format=-2.1] S[table-format=-2.1] S[table-format=-2.1] S[table-format=-2.1]}
                        \toprule
                        False Positive          & {0.01\%}  & {0.1\%}   & {1\%}     & {10\%}     & {25\%}     \\
                        \midrule
                        $\alpha_{E} = 0.25$     & 0.72  & 5.61 & 22.01 & 58.38 & 84.20  \\
                        $\alpha_{E} = 0.50$     & 0.71  & 8.66 & 24.39 & 66.07 & 87.76  \\
                        $\alpha_{E} = 0.75$     & 0.16  & 4.69 & 17.60 & 67.85 & 89.12  \\
                        $\alpha_{E} = 1.00$     & 2.69  & 8.53 & 24.64 & 74.24 & 93.07  \\
                        \bottomrule
                    \end{tabular}
                \end{center}
                \label{tab:flickr_pert}
            \end{table}

        \subsection{Error Analysis}\label{sec:analysisoferror}
            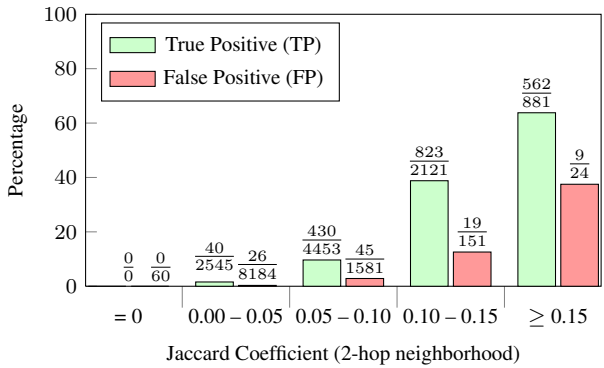
\begin{figure}[h!]
                \pgfplotstableread{
                0 0        0
                1 1.57     0.31
                2 9.66     2.85
                3 38.8     12.58
                4 63.79    37.5
                }\dataset
                \begin{tikzpicture}
                    \begin{axis}[ybar,
                        width=\linewidth,
                        height=5.2cm,
                        compat=newest,
                        bar width=5mm,
                        ymin=0,
                        ymax=100,
                        xlabel={Jaccard Coefficient (2-hop neighborhood)},
                        ylabel={Percentage},
                        xtick=data,
                        xtick pos=left,
                        ytick pos=left,
                        font=\small,
                        xticklabels = {
                            \strut = 0,
                            \strut 0.00 -- 0.05,
                            \strut 0.05 -- 0.10,
                            \strut 0.10 -- 0.15,
                            \strut $\geq$ 0.15
                        },
                        major x tick style = {opacity=0},
                        minor x tick num = 1,
                        minor tick length=2ex,
                        every node near coord/.append style={
                                anchor=west,
                                rotate=90
                        },
                        legend entries={True Positive (TP), False Positive (FP)},
                        legend columns=1,
                        legend pos=north west,
                        legend style={cells={anchor=center, fill},nodes={inner sep=3pt}},area legend
                        ]
                        \addplot[draw=black, fill=green!20, postaction={pattern=north east lines}] table[x index=0,y index=1]\dataset;
                        \addplot[draw=black, fill=red!40, postaction={pattern=dots}] table[x index=0,y index=2] \dataset;

                        \node at (axis cs:0,7) [anchor=center, font=\normalsize] {$\frac{0}{0}$};
                        \node at (axis cs:0.3,7) [anchor=center, font=\normalsize] {$\frac{0}{60}$};

                        \node at (axis cs:0.8,11) [anchor=center, font=\normalsize] {$\frac{40}{2545}$};
                        \node at (axis cs:1.2,8) [anchor=center, font=\normalsize] {$\frac{26}{8184}$};

                        \node at (axis cs:1.8,17) [anchor=center, font=\normalsize] {$\frac{430}{4453}$};
                        \node at (axis cs:2.2,10) [anchor=center, font=\normalsize] {$\frac{45}{1581}$};

                        \node at (axis cs:2.8,46) [anchor=center, font=\normalsize] {$\frac{823}{2121}$};
                        \node at (axis cs:3.2,20) [anchor=center, font=\normalsize] {$\frac{19}{151}$};

                        \node at (axis cs:3.8,71) [anchor=center, font=\normalsize] {$\frac{562}{881}$};
                        \node at (axis cs:4.2,45) [anchor=center, font=\normalsize] {$\frac{9}{24}$};
                    \end{axis}
                \end{tikzpicture}
                \caption{True Positive and False Positive rates vary widely depending on the social neighborhood overlap}
                \label{fig:analysisoferror}
            \end{figure}

            This section investigates a key source of false positives, namely graph overlap between distinct nodes. The features used to characterize and match nodes are based on the degrees of its neighbors. Therefore we expect nodes that share a significant number of friends would be harder to distinguish than those sharing few.

            To study this hypothesis, we classify 10\,000 identical pairs and 10\,000 non-identical pairs, that have been anonymized using the edge perturbation technique
            (with $\alpha_E = 0.25$, $50$ seeds, see Section~\ref{sec:traditional}). The classifier is tuned to achieve an overall 1\% false positive rate, leading to an overall true positive rate of about 20\%. Pairs are also categorized according to the Jaccard Coefficient (JC) of their 2-hop social networks, on which their feature vectors are computed, which provides a degree of social overlap.

            Figure~\ref{fig:analysisoferror} illustrates that the True Positive and False Positive rates vary widely depending on the social overlap of the tested pairs. When the overlap is small (JC 0.00 -- 0.05) the false positive rate is extremely low at 0.31\%, but the true positive rate also suffers greatly. However, when the overlap is significant (JC $\geq$ 0.15) both True Positive (63.79\%) and False Positive rates (37.5\%) rise significantly above the overall baseline. However, only a very small fraction of non-identical pairs have such a high JC (0.24\%) , compared to a large fraction of identical pairs (8.81\%), leading to a small overall error for this category.

            We conclude that False Positives links are likely to occur with nodes in the social vicinity of the actual match. Thus, even erroneous positives allow an adversary to identify the social neighborhood of a match, even if the exact matched node is incorrect.


    \section{Discussion}\label{sec:Discussion}
        We have shown that the machine learning approach to de-anonymization and linkage can be successful. Furthermore, we establish that training can be performed on different graphs than the ones being attacked. Our core evaluation is performed by training on a small separate set of egonets derived from the same distribution for training and testing. The resulting classifiers are able to de-anonymize egonets with no previously seen nodes. Additionally, classifiers trained on data from a totally different distribution (Epinions and Pokec cross-classification) still perform well enough to classify a non-negligible fraction of pairs as identical, even for low false positive rates. At the very least they can be used to identify a number of common seed nodes, to support further linkage that can tolerate higher false positive rates. We also show that the classifier generalizes well and successfully handles de-anonymization task in the traditional setting. This provides further evidence of its applicability to evaluate novel anonymization schemes applied to diverse datasets. We expect this type of analysis to be most relevant to the evaluation of any new anonymization algorithm, since as demonstrated finding suitable graphs and generating training data is easy.

        \para{Resilience of classifier.} The resilience of the classifiers is due to the differential nature of the training and the classification task: the training algorithm is provided pairs of nodes, resulting from an anonymization algorithm, and provided labels about whether they are the same or not. As a result it learns invariants that are characteristic of the anonymization method, not merely the data. One might na\"{i}vely conjecture that basing the node feature vectors on neighborhood degree distribution would be inept in attacking anonymization schemes that perturb the node degree at random (Section \ref{sec:traditional}). The success of classification is not conditioned on the ``invariance'' of the node degrees but on their ``variance'' as a function of the anonymization strategy. This function can be learned and hence hedged to attack the scheme. We conjecture that purely synthetic specially crafted graphs may be equally (or even better) suited to train the classifiers, but we leave this investigation for future work.

        \subsection{Is Anonymization Effective?}
            So are anonymization Schemes~1 \& 2 effective? First, it is clear that Scheme~2 is mildly more effective than Scheme~1. Given an egonet, Scheme~1 exposes more nodes to our attack than Scheme~2 due to lesser damage to the node degrees. Also, the true positive rate of the classifier is lower for any fixed acceptable false negative rate. However, even for extremely low false positive rates of $0.01\%$ a non-negligible fraction of the nodes are correctly classified ranging from $1.86\%$ for 2-hop nodes to $35.08\%$ for 1-hop nodes (Epinions) and from $1.62\%$ to $4.20\%$ (Pokec). First, a person may with non-negligible probability be within a successfully linked pair. Second, extremely high confidence matches can be used to piece together disparate egonets into larger graphs. Those larger graphs, with some common nodes, can be then further used to de-anonymize other users.

            Therefore we believe that the linking rates we observed are too high for the original social network to be effectively unrecoverable. We conclude that the D4D competition organizers were prudent to limit the disclosure of the dataset to known participants and require contractual assurances that they will not de-anonymize the data.

        \subsection{Improving De-anonymization}
            The D4D data release provides access to egonets observed between different time slots, with same identifiers across time slots (see Section~\ref{sec:data release}). We can amplify our classification success by classifying a candidate node pair in each time slot and then applying a majority rule for deciding the true association -- identical or non-identical. Such an attack is significantly more potent than one possible on an aggregated social graph across all time periods. The original D4D dataset also contained edge weights. We chose to ignore those to devise de-anonymization strategies for generic graphs without such weights. However, the feature vector proposed could be augmented to take those weights into account to generate distinct features.

            A key difference between the proposed de-anonymization algorithm, and previous work~\cite{narayanan2009anonymizing}, is the lack of reliance on known ``seeds''. Those seeds are adversary side-information, i.e.\ a few known nodes between two unlinkable networks, that can be used to unravel the anonymization. Our algorithms take two totally distinct egonets, for which no common node is known, and classify nodes within them as identical or non-identical. In case some seeds are known, our techniques can be applied to the common neighbors of the seeds to determine whether they are the same node or not. In that case larger false positive rates can be tolerated since the a-priori probability of the nodes being the same is larger (by many orders of magnitude) compared to any two random nodes. Since the proposed approach works extremely well for larger false positive rates (like $1\%$) we expect such a combined scheme to be extremely effective. A full investigation of the iterative application of the presented method, to build up a full graph is beyond the scope of this work.

        \subsection{Choosing Tree Parameters}\label{sec:param}
            In this section we discuss the choice of parameters to train decision trees.

            \noindent {\bf Forest size.} Literature suggests~\cite{criminisi2011regression,shotton2008semantic,yin2007tree} that testing accuracy increases monotonically with the forest size $T$. Using a small forest size decreases the accuracy of uncertainty and produces low quality confidence estimates, this leads to erroneous generalization. Criminisi et al. present examples in their report~\cite{Criminisi2011TR} which obtain good results for $T=400$, hence we have chosen the forest size to be 400. We tested our results with $T=500$ without any appreciable gain.

            \noindent {\bf Randomness.} We inject randomness while training using bagging and random node optimization, which leads to a lower confidence in the posterior, but smoother and more spread out posteriors. We use bagging by passing 600 node pairs (per class) sampled at random from the training data set (Table \ref{tab:sample_epi_pok}, Appendix \ref{appendix:Sample Sizes}) to each tree. Every split node is exposed to 5\% of the total feature parameters. We experimented by increasing the percentage to 10\%, 20\% and 25\% but did not observe any benefits.

            \noindent {\bf Features.} We use the feature vector length as 70 and bin size as 15. The intuition behind this choice is that most individuals in a social network have fewer than 1\,000 friends and the above choice reflects this. We study the impact of vector length and bin sizes by training a random forest with $T=200$ and bagging with 200 node pairs (per class) for Epinions Scheme~2 (complete set of node pairs). We experimented with vector lengths 21, 35, 105 and bin sizes 50, 30, 10 respectively. Increasing the vector length increases the complexity by a quadratic factor. Table \ref{tab:epi_bucket} and Figure \ref{fig:roc_epi_vec_len} (Appendix \ref{appendix:ROC}) demonstrates that the performance is similar for different vector lengths. The vector length of 21 performs the best for the demonstrated example but we found that our choice of vector length 70 works best over all across multiple settings.

            Finally, we experimented with D4D by selecting features beyond the 1-hop neighborhood of the node pair under consideration. This sharply increased our false positive rates. This is due to the fact that many nodes share large parts of their 2-hop neighborhood, and the number of features in that space is much larger than than in the 1-hop neighborhood. This is in contrast with the de-anonymization features used in Section \ref{sec:traditional}. This happens because the global properties of the overlapping sub-graph are better preserved when the full graph is available instead of just the local neighborhood. Thus allowing us to select the features from the diverse 2-hop network without sharp increase in false positives. Picking parameters which are optimal across all settings is a hard task, instead we focus on developing techniques which perform well for appropriately chosen parameters.

            \begin{table}[h!]
                \caption{Epinions: Varying vector length (Scheme~2 - Complete): \emph{False Positive} vs. \emph{True Positive}}
                \begin{center}
                    \textbf{Varying feature vector length}\\
                    \begin{tabular}{ l S[table-format=-2.1] S[table-format=-2.1] S[table-format=-2.1] S[table-format=-2.1] S[table-format=-2.1]}
                        \toprule
                        False Positive   & {0.01\%}  & {0.1\%}  & {1\%}  & {10\%}  & {25\%}      \\
                        \midrule
                        21 X 50          & 1.07  & 11.16 & 37.62 & 90.88 & 97.35  \\
                        35 X 30          & 0.40  & 3.78  & 22.97 & 91.45 & 97.75  \\
                        70 X 15          & 0.49  & 1.49  & 21.01 & 91.87 & 97.99  \\
                        105 X 10         & 0.49  & 2.39  & 21.38 & 92.37 & 97.82  \\
                        \bottomrule
                    \end{tabular}
                \end{center}
                \label{tab:epi_bucket}
            \end{table}


    \section{Related Work}\label{sec:Related Work}
        Anonymizing social networks has proven to be a tough challenge. Backstrom et al. presented~\cite{backstrom2007wherefore} active (using sybil nodes) and passive attacks based on searching patterns in sub-graphs to learn relationships between pre-selected target individuals of an anonymised social graph. Group membership has been shown~\cite{wondracek2010practical} to be sufficient to identify an individual in a social network.

        Narayanan and Shmatikov present~\cite{narayanan2009anonymizing} a de-anonymization attack on a social network using auxiliary information from a different social network. They are also the first to note that a large volume of matching errors will be in the vicinity of the actual matching nodes. Similar techniques have been used to attack Netflix dataset by co-relating it with IMDb dataset~\cite{narayanan2008robust}. We test our algorithm in this setting but also highlight some key differences that makes their approach hard to apply to the D4D dataset. The algorithm proposed aims to link two large correlated social networks. However, the D4D dataset is split into small ego-nets. Consequently, one cannot straight-forwardly apply a method based on identifying few known seed users and expanding from them, since such a propagation would naturally end at the boundaries of each ego-net. For this reason our learning approach does not make use of known seeds, but instead we use training examples that do not necessarily comprise the nodes in the sub-graph to be de-anonymized. This allows the proposed approach to train on different graphs, and still link nodes in small subgraphs of other networks. However, we note that when a large subgraph is available the two approaches may be combined: one may use the techniques proposed in this paper to identify few seeds with high certainty, and then apply our techniques to identify known nodes in their vicinity, which are more likely to be related than random nodes.

        Narayanan et al. de-anonymize~\cite{narayanan2011link} the Kaggle dataset released for link prediction using a pre-crawled version of the same dataset. Their work combines de-anonymization and link prediction using random forests, however, the de-anonymization phase does not use any machine learning. Random forests are used to predict those links which pure de-anonymization could not. Additionally, their work is based on availability of ground truth to mount de-anonymization attack on a dataset which is not adversarial. Cukierski et al. got very good results\footnote{\begin{scriptsize}\url{http://blog.kaggle.com/2011/01/14/how-i-did-it-will-cukierski-on-finishing-second-in-the-ijcnn-social-network-challenge/}\end{scriptsize}} just using pure random forests for link prediction, rather than de-anonymization, for the same dataset. In contrast, we do not have directionality available for our graphs and our feature extraction is simple and efficient, an important factor for attacking huge datasets.

        Our work uses these techniques for the first time in an adversarial de-anonymization setting: whereas previous de-anonymization techniques considered graphs that were organically noisy but structurally intact. In contrast the D4D challenge organizers purposefully and aggressively alter graph topology to prevent linkage.


    \section{Conclusion}\label{sec:Conclusion}
        Privacy leaks could have serious consequences when they pertain to countries like Ivory Coast which has a history of civil wars and political unrest. Thus carefully evaluating the privacy of data anonymization schemes, such as the ones made available by the D4D competition is imperative. Traditionally such an analysis had to be performed manually, and painstakingly repeated for any new variant of the anonymization scheme, despite general results indicating social network anonymization schemes are likely to be broken. Our approach bypasses the need for manual effort, and can uncover artefacts of the anonymization process just using examples that allow to re-linking nodes in ``anonymized'' networks.

        In terms of the specific anonymization procedures for the D4D competition motivating this study: we do conclude that Scheme~2 is marginally harder to de-anonymize and re-link than the original Scheme~1 we evaluated. Despite this, neither provide a level of privacy we would recommend for public release of data. Even a reliable linkage rate of 1\% would lead to a significant fraction of the mobile operator customers being potentially linked. Big data makes even a relatively low probability linkage events a certainty. In fact for a number of realistic configurations we show the linkage rate was much higher, leading to efficient attacks particularly if some side information is available to build good priors and tolerate slightly higher false positives. Thus we are relieved that the competition only released such data under confidentiality agreements.

        The approach, and learning task we rely upon, to link or de-anonymize, are purposefully simple: it only uses the graph topology, not attributes, directionality or multiple snapshots of graphs over time, despite the availability of such data in some cases. It also considers a single pass of the procedure, where no known seeds are available. It is clear that richer features could be used, and the attack can be iterated once a handful of nodes have been uncovered to unravel larger graphs. These could be topics for future work, and perfecting such attacks could be a fruitful research field for years.

        However, pursuing too far such a research program may yield diminishing returns. Our results confirm previous wisdom that releasing anonymized social networks is likely to, either result in privacy catastrophe, or very poor utility. This paper demonstrates that even an automated algorithm can produce \emph{good enough} de-anonymization attacks. Thus, it may be more beneficial to research instead alternative ways to perform social network analysis in a privacy friendly manner, and without the need for ``anonymized'' graphs.


        \para{Acknowledgments.} We thank Professor Vincent Blondel and Orange for giving us the opportunity to study the anonymiation strategy for D4D. The authors would like to thank Pushmeet Kohli and Jamie Shotton for advice on machine learning, as well as Rebecca Murphy, Richard Clayton and Steven Murdoch for comments on the draft manuscripts. This work was supported by the Engineering and Physical Sciences Research Council [grant number EP/J500665/1]; and Microsoft Research through its PhD Scholarship Programme.


    \bibliographystyle{abbrv}
    \bibliography{d4d}

    \appendix
    \section{Definitions}\label{appendix:Definitions}
        \begin{enumerate}
            \item Ego - A graph node around which the egonet is generated.

            \item Egonet - The local network centered around an ego derived based on some function.

            \item $n$-hop node - A node such that the shortest path length from the node to the ego is $n$.

            \item $n$-hop network - A node induced neighborhood graph around an ego with all $n$-hop nodes included. It is also known as $n$-hop neighborhood.
        \end{enumerate}

    \section{Data Sample Sizes} \label{appendix:Sample Sizes}
        To run cross classification corresponding sample from the complimentary dataset is used. For example to run cross classification for Epinions Scheme~1 for 1-hop node pairs we used 451 Scheme~1 1-hop node pairs sampled from Pokec.

        \begin{table}[h!]
            \caption{Training and testing number of pairs for Epinions and Pokec}
            \begin{center}
                \parbox{\linewidth}{
                \centering
                \begin{tabular}{ l S[table-format=-5] S[table-format=-4] S[table-format=-5] S[table-format=-4] }
                    {\bf Epinions} & \multicolumn{2}{c}{\textbf{Scheme~1}} & \multicolumn{2}{c}{\textbf{Scheme~2}} \\

                    \toprule
                    ~                   & {Train}      & {Test} & {Train}      & {Test} \\
                    \midrule
                    1-hop\footnotemark  & 579          & 812   & 959          & 975   \\
                    1,2-hop             & 51716        & 10000 & 11407        & 10000 \\
                    2-hop               & 1910868      & 10000 & 46830        & 10000 \\
                    Complete            & 1963163      & 10000 & 59196        & 10000 \\
                    Non-identical       & 48910        & 10000 & 35655        & 10000 \\
                    \bottomrule
                \end{tabular}
                }
                \parbox{\linewidth}{
                \vspace{3mm}
                \centering
                \begin{tabular}{ l S[table-format=-5] S[table-format=-4] S[table-format=-5] S[table-format=-4]}
                    {\bf Pokec} & \multicolumn{2}{c}{\textbf{Scheme~1}} & \multicolumn{2}{c}{\textbf{Scheme~2}} \\

                    \toprule
                    ~                                     & {Train}      & {Test} & {Train}      & {Test} \\
                    \midrule
                    1-hop\footnotemark[\value{footnote}]  & 451          & 480    & 3226         & 3075  \\
                    1,2-hop                               & 39529        & 10000  & 8418         & 7210  \\
                    2-hop                                 & 1036966      & 10000  & 9263         & 7349  \\
                    Complete                              & 1145419      & 10000  & 20907        & 10000 \\
                    Non-identical                         & 124171       & 10000  & 495353       & 10000 \\
                    \bottomrule
                \end{tabular}
                }
            \end{center}
            \label{tab:sample_epi_pok}
        \end{table}

        \footnotetext{\begin{scriptsize}When there were fewer than 600 node pairs we trained each tree with all the nodes available.\end{scriptsize}}

        \begin{table}[h!]
            \caption{Scheme~1: Sample sizes for ad-hoc de-anonymization of Case~1 node pairs}
            \centering
            \textbf{Sample sizes}\\
            \begin{tabular}{ l S[table-format=-4.2] S[table-format=-4.2]}
                \toprule
                ~                & {Identical}   & {Non-identical} \\
                \midrule
                Epinions         & 900           & 4907  \\
                Pokec            & 1000          & 5000  \\
                \bottomrule
            \end{tabular}
            \label{tab:adhocsize}
        \end{table}

        \begin{table}[h!]
            \caption{Training and testing number of pairs for Flickr}
            \parbox{\linewidth}{
            \centering
            \textbf{Varying number of seeds ($\alpha_E = 0.25$)}\\
            \begin{tabular}{ l S[table-format=-4.2] S[table-format=-4.2] }
                \toprule
                ~                   & {Train}      & {Test} \\
                \midrule
                10 seeds            & 10           & 10000 \\
                50 seeds            & 50           & 10000 \\
                250 seeds           & 250          & 10000 \\
                1250 seeds          & 1250         & 10000 \\
                Non-identical       & 5000         & 10000 \\
                \bottomrule
            \end{tabular}
            }
            \parbox{\linewidth}{
            \vspace{3mm}
            \centering
            \textbf{Varying edge overlap (50 seeds)}\\
            \begin{tabular}{ l S[table-format=-4.2] S[table-format=-4.2] }
                \toprule
                ~                       & {Train}      & {Test} \\
                \midrule
                $\alpha_E = 0.25$       & 50           & 10000 \\
                $\alpha_E = 0.50$       & 50           & 10000 \\
                $\alpha_E = 0.75$       & 50           & 10000 \\
                $\alpha_E = 1.00$       & 50           & 10000 \\
                Non-identical           & 5000         & 10000 \\
                \bottomrule
            \end{tabular}
            }
            \label{tab:sample_flickr}
        \end{table}

        \section{Performance}\label{appendix:Performance}
            All experiments were run on a commodity laptop. Code for the entire project is written in pure Python and run using CPython. Time taken to extract features as described in Section~\ref{sec:features} is less than 6 minutes for Scheme~1 and less than a minute for Scheme~2 for both datasets. Table \ref{tab:perf_epi_pok} shows the time taken to train forests with 400 trees of each class of identical node pairs and run the classifier for both schemes and datasets. Testing time of cross classification is similar to the corresponding reported times. Since trees in a forest are independent of each other, training and testing can be run in parallel. However we present the time it takes to perform these operations on a single core. Table \ref{tab:perf_flickr} includes the same details for Flickr edge perturbation strategy.

            \begin{table}[h!]
                \caption{Training and testing times for Epinions and Pokec}
                \begin{center}
                    \parbox{\linewidth}{
                    \centering
                    \begin{tabular}{ l S[table-format=-3.2] S[table-format=-3.2] S[table-format=-3.2] S[table-format=-3.2]}
                        {\bf Epinions} & \multicolumn{2}{c}{\textbf{Scheme~1}} & \multicolumn{2}{c}{\textbf{Scheme~2}} \\
                        \toprule
                        Time (hrs)          & {Train}      & {Test} & {Train}      & {Test} \\
                        \midrule
                        1-hop               & 14.93        & 0.86  & 17.94        & 1.15  \\
                        1,2-hop             & 15.96        & 1.60  & 15.93        & 2.09  \\
                        2-hop               & 15.52        & 1.87  & 15.44        & 1.59  \\
                        Complete            & 15.58        & 1.96  & 16.17        & 2.08  \\
                        \bottomrule
                    \end{tabular}
                    }
                    \parbox{\linewidth}{
                    \vspace{3mm}
                    \centering
                    \begin{tabular}{ l S[table-format=-3.2] S[table-format=-3.2] S[table-format=-3.2] S[table-format=-3.2]}
                        {\bf Pokec} & \multicolumn{2}{c}{\textbf{Scheme~1}} & \multicolumn{2}{c}{\textbf{Scheme~2}} \\
                        \toprule
                        Time (hrs)          & {Train}      & {Test} & {Train}      & {Test} \\
                        \midrule
                        1-hop               & 10.94        & 1.00  &20.04         &1.37   \\
                        1,2-hop             & 14.58        & 2.18  &18.15         &2.05   \\
                        2-hop               & 17.07        & 2.35  &14.56         &1.43   \\
                        Complete            & 16.80        & 2.50  &18.40         &2.36   \\
                        \bottomrule
                    \end{tabular}
                    }
                \end{center}
                \label{tab:perf_epi_pok}
            \end{table}

            \begin{table}[h]
                \caption{Training and testing times for Flickr}
                \parbox{\linewidth}{
                \centering
                \textbf{Varying number of seeds ($\alpha_E = 0.25$)}\\
                \begin{tabular}{ l S[table-format=-4.2] S[table-format=-4.2] }
                    \toprule
                    Time (hrs)          & {Train}      & {Test} \\
                    \midrule
                    10 seeds            & 0.72         & 0.27 \\
                    50 seeds            & 3.21         & 0.79 \\
                    250 seeds           & 5.42         & 0.97 \\
                    1250 seeds          & 14.94        & 1.28 \\
                    \bottomrule
                \end{tabular}
                }

                \parbox{\linewidth}{
                \vspace{3mm}
                \centering
                \textbf{Varying edge overlap (50 seeds)}\\
                \begin{tabular}{ l S[table-format=-4.2] S[table-format=-4.2] }
                    \toprule
                    Time (hrs)              & {Train}      & {Test} \\
                    \midrule
                    $\alpha_E = 0.25$       & 3.21         & 0.79 \\
                    $\alpha_E = 0.50$       & 2.89         & 0.64 \\
                    $\alpha_E = 0.75$       & 3.07         & 0.66 \\
                    $\alpha_E = 1.00$       & 2.88         & 0.68 \\
                    \bottomrule
                \end{tabular}
                }
                \label{tab:perf_flickr}
            \end{table}

    \section{ROC curves for classification}\label{appendix:ROC}
        The diagonal line denotes the FP vs TP akin to guessing.
        \begin{figure}[h!]
            \centering
            \begin{subfigure}{.5\linewidth}
                \centering
                {\includegraphics[width=1.05\linewidth]{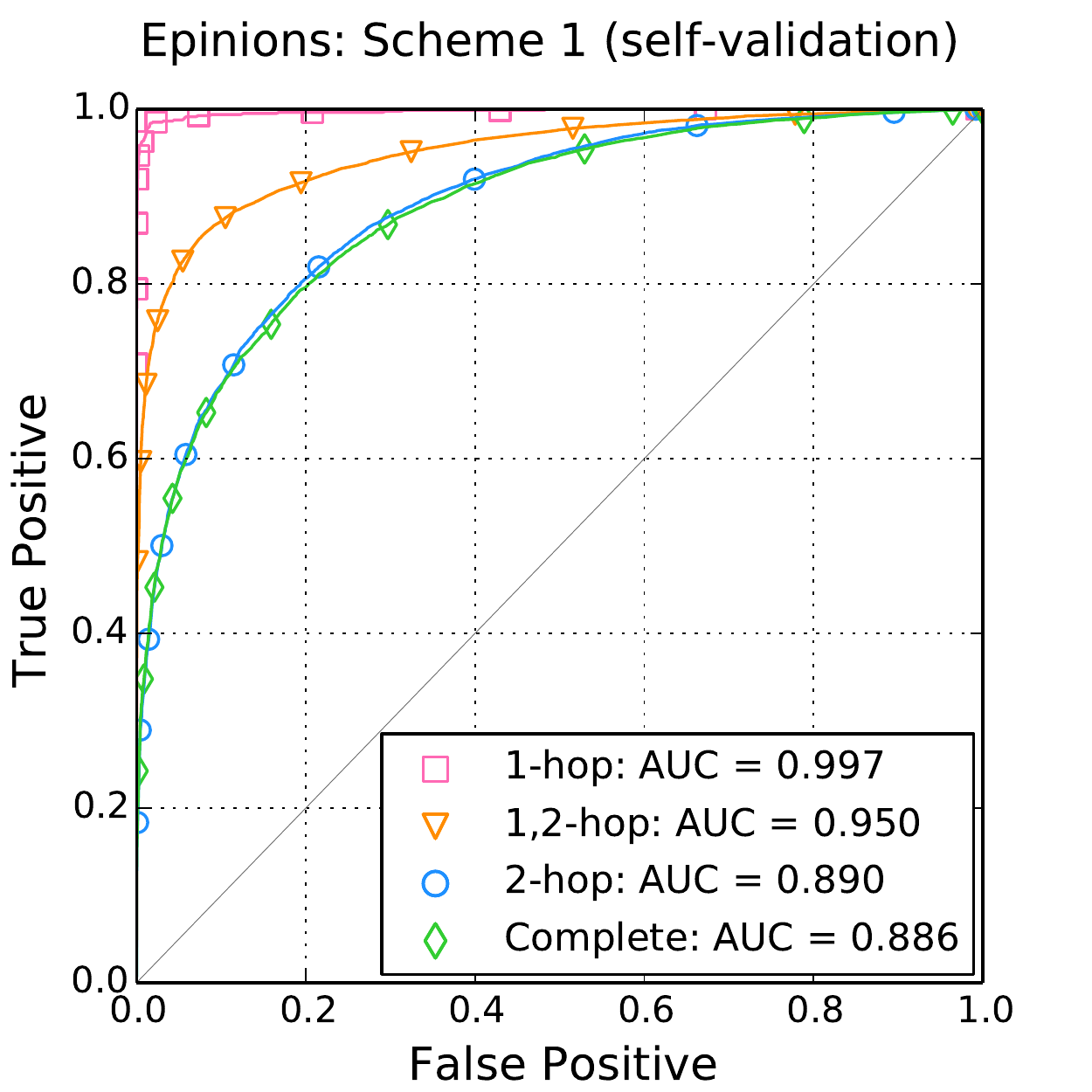}}
            \end{subfigure}%
            \begin{subfigure}{.5\linewidth}
                \centering
                {\includegraphics[width=1.05\linewidth]{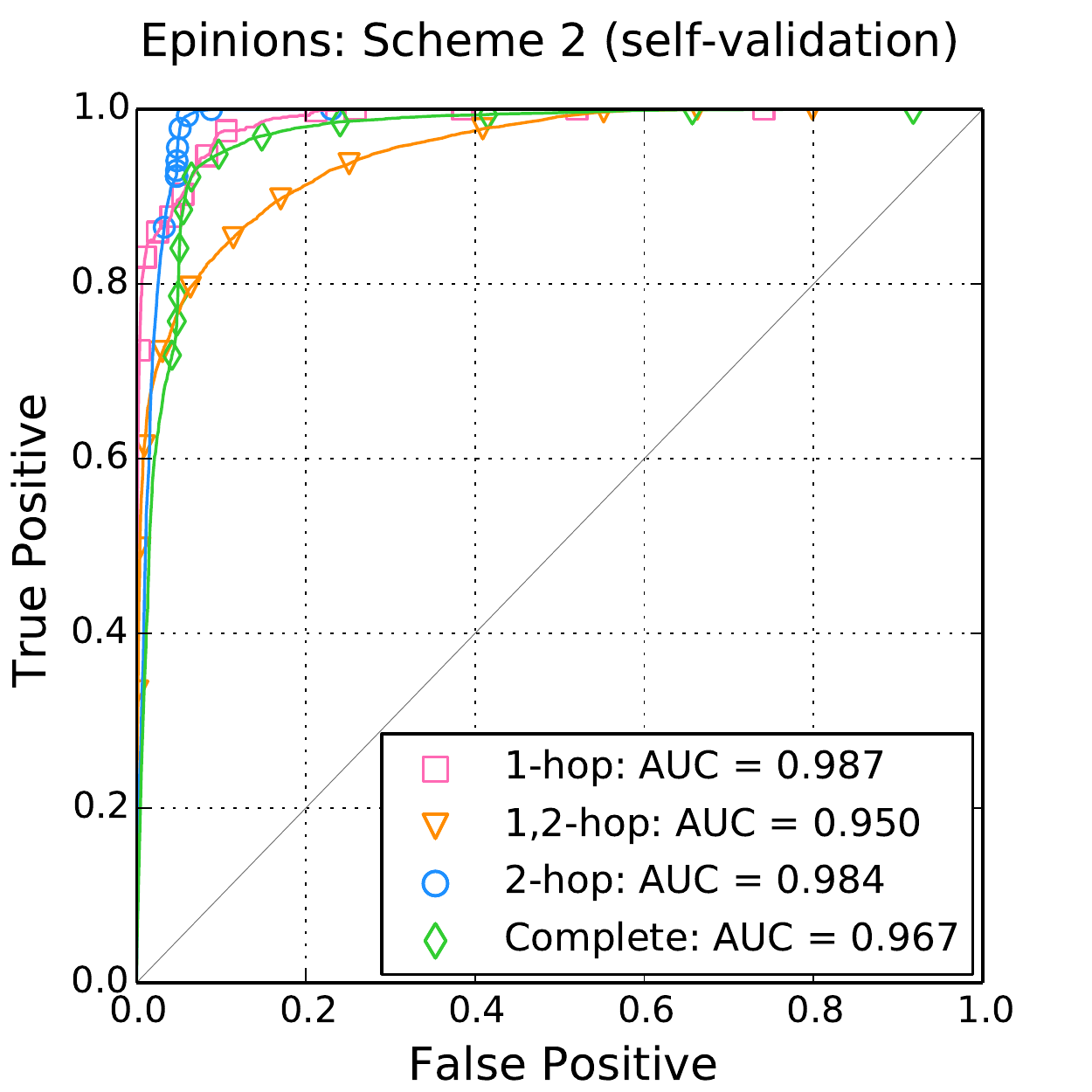}}
            \end{subfigure}%
            \caption{Epinions (self-validation): ROC curves for both schemes}
            \label{fig:roc_epi_self}
        \end{figure}

        \begin{figure}[h!]
            \centering
            \begin{subfigure}{.5\linewidth}
                \centering
                {\includegraphics[width=1.05\linewidth]{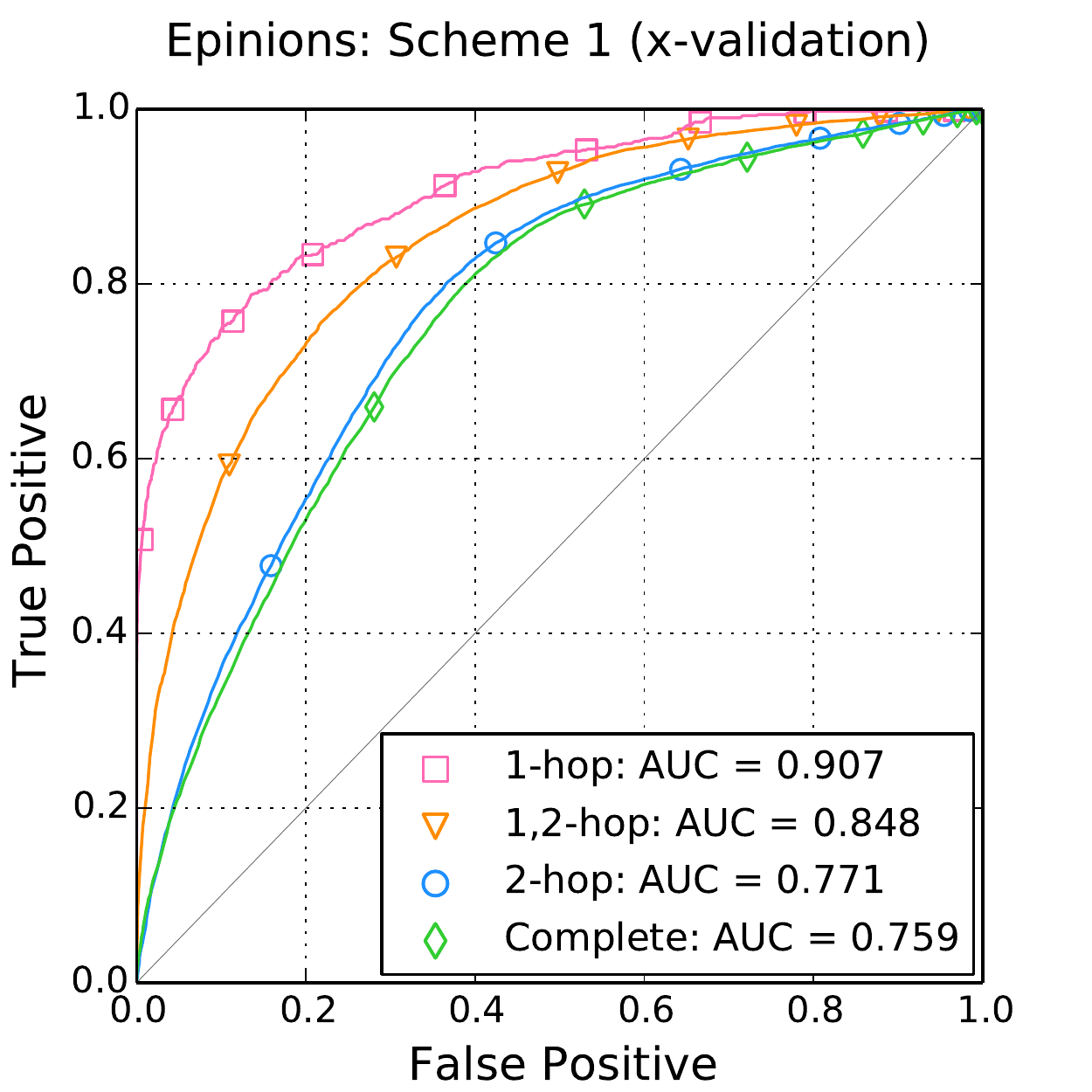}}
            \end{subfigure}%
            \begin{subfigure}{.5\linewidth}
                \centering
                {\includegraphics[width=1.05\linewidth]{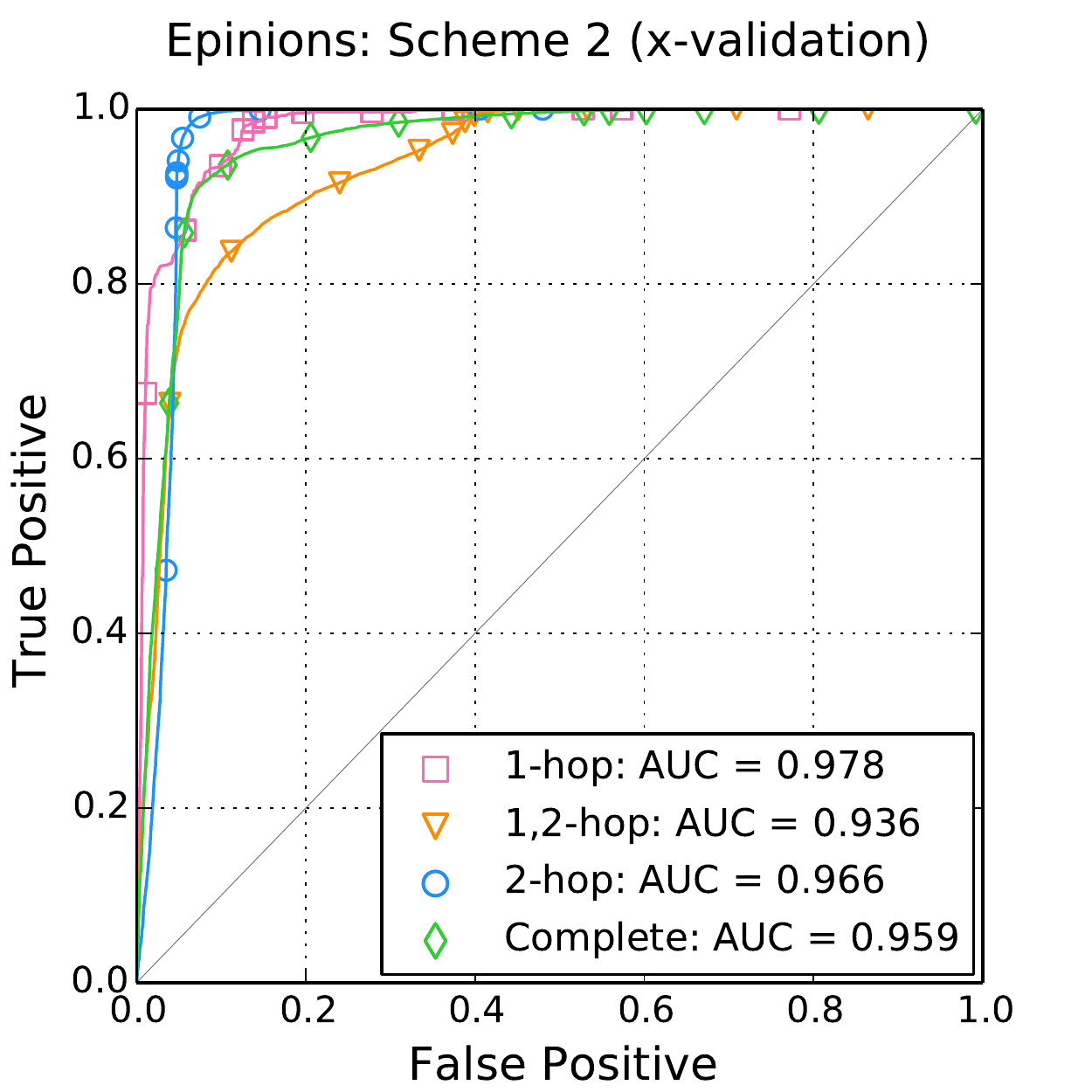}}
            \end{subfigure}%
            \caption{Epinions (x-validation): ROC curves for both schemes}
            \label{fig:roc_epi_x}
        \end{figure}

        \begin{figure}[h!]
            \centering
            \begin{subfigure}{.5\linewidth}
                \centering
                {\includegraphics[width=1.05\linewidth]{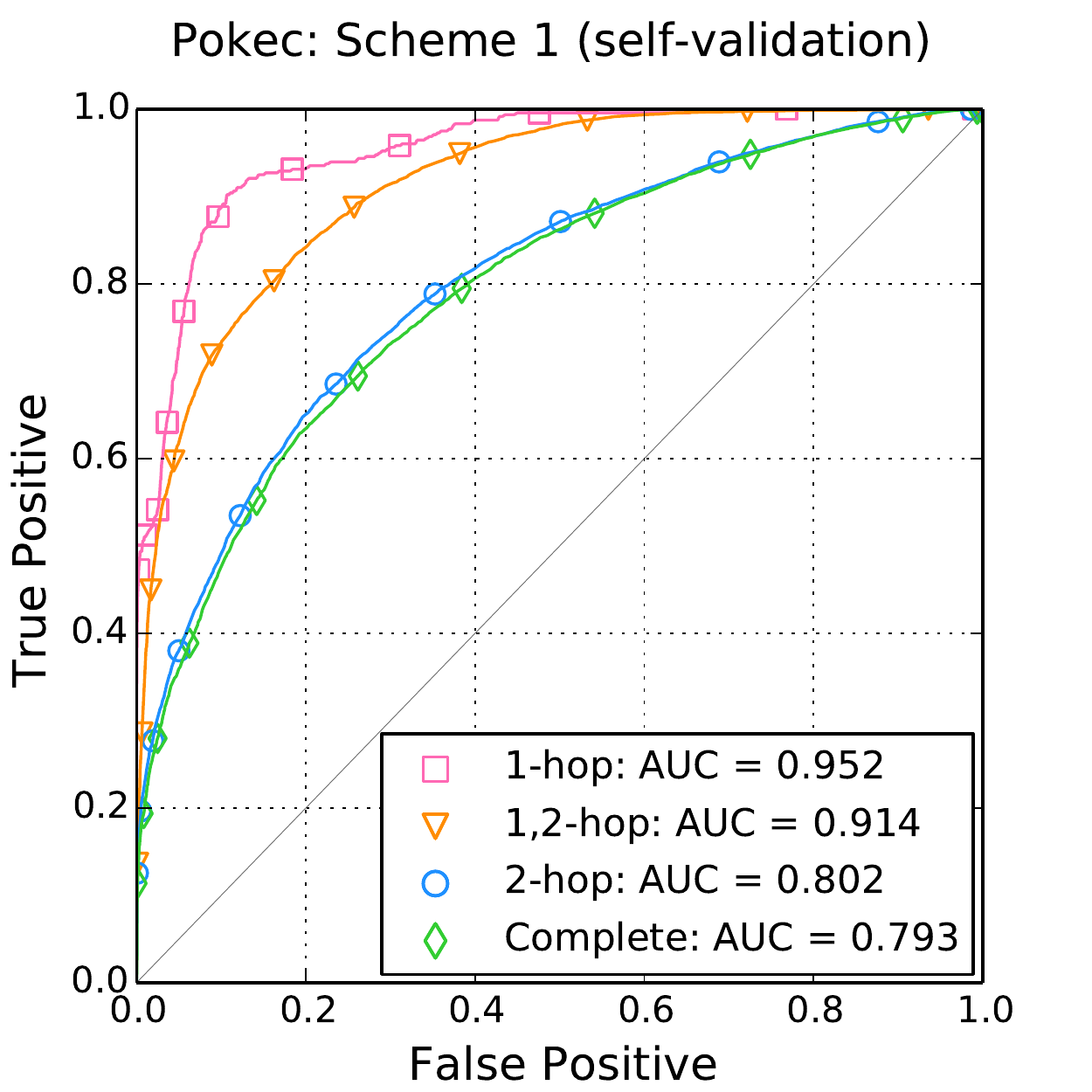}}
            \end{subfigure}%
            \begin{subfigure}{.5\linewidth}
                \centering
                {\includegraphics[width=1.05\linewidth]{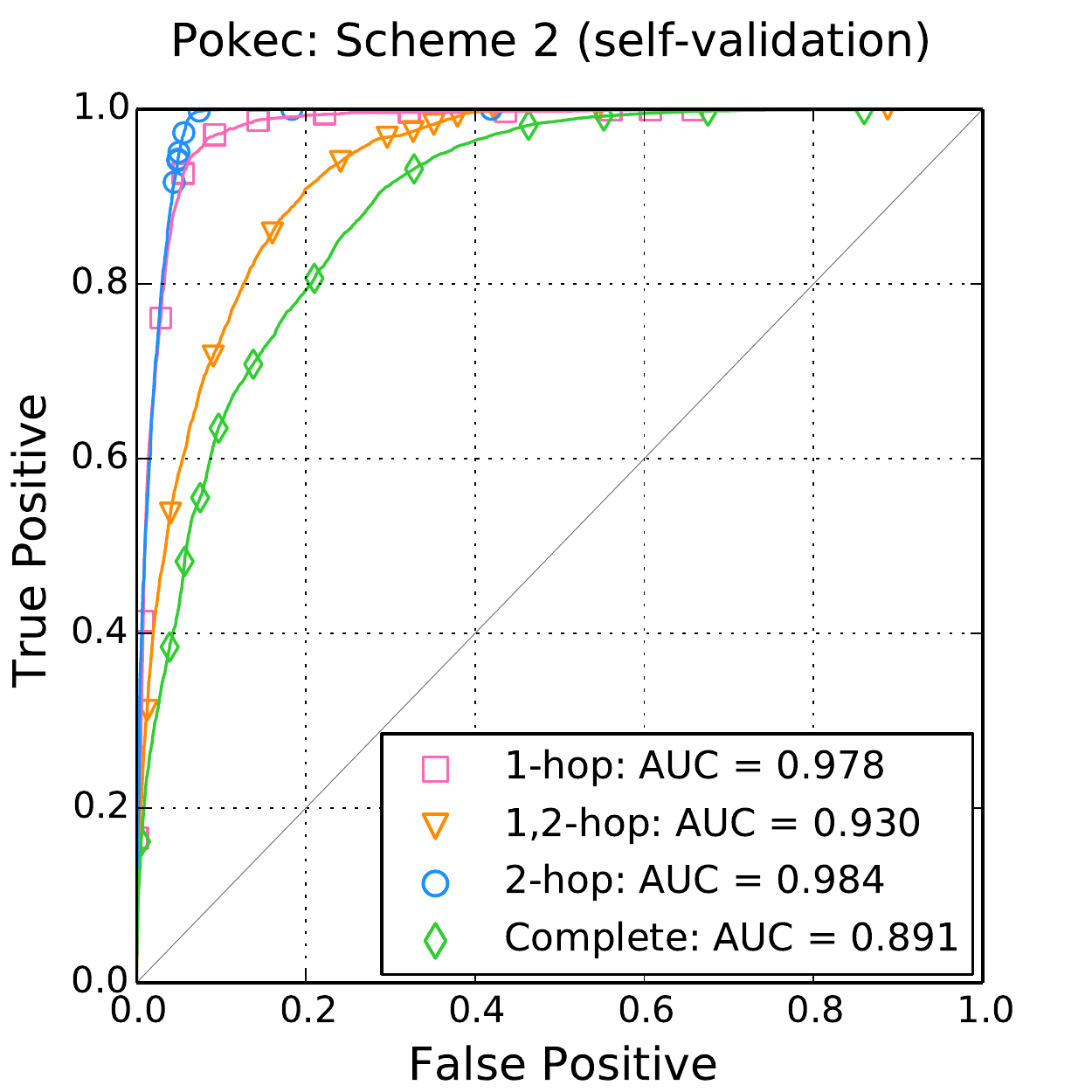}}
            \end{subfigure}%
            \caption{Pokec (self-validation): ROC curves for both schemes}
            \label{fig:roc_pok_self}
        \end{figure}

        \begin{figure}[h!]
            \centering
            \begin{subfigure}{.5\linewidth}
                \centering
                {\includegraphics[width=1.05\linewidth]{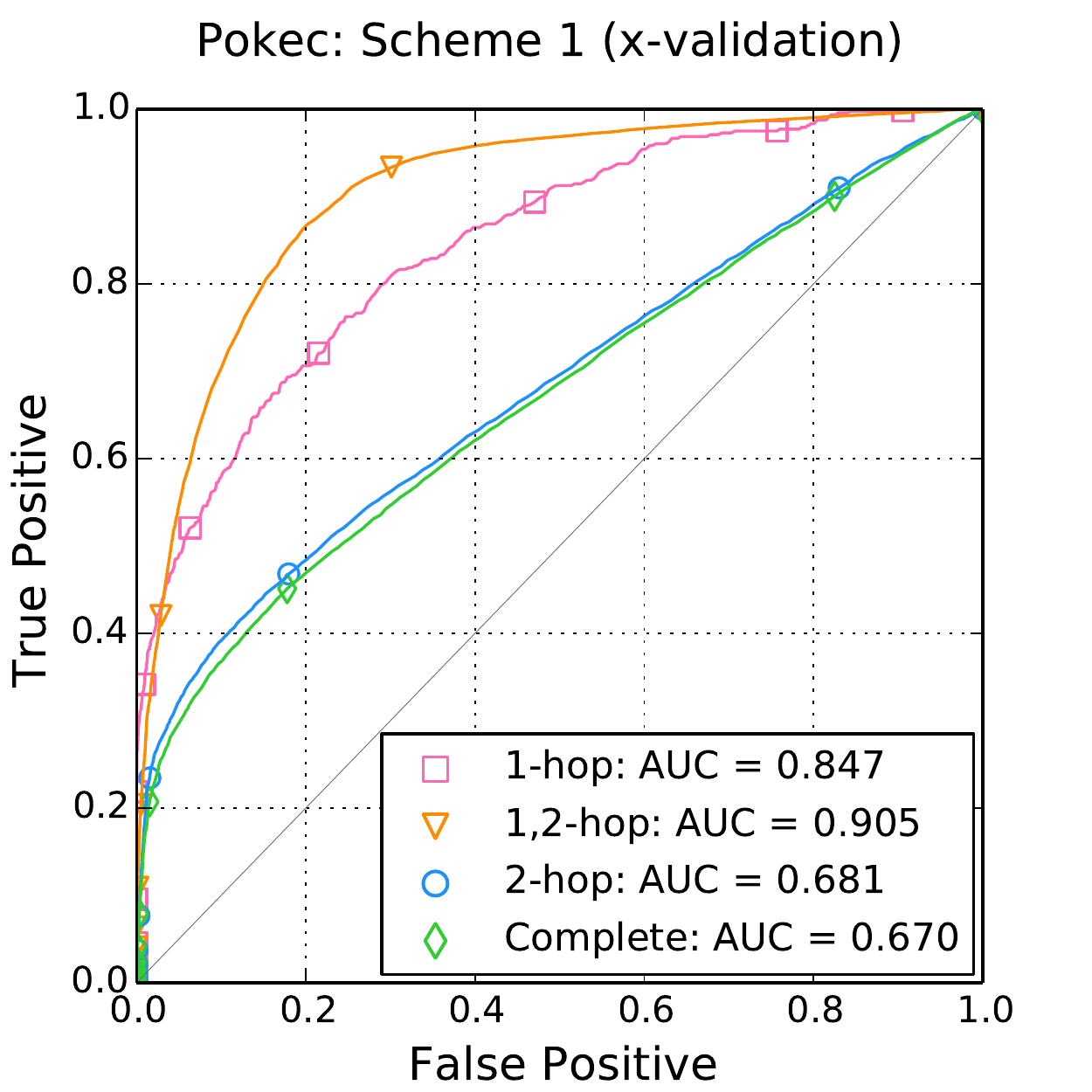}}
            \end{subfigure}%
            \begin{subfigure}{.5\linewidth}
                \centering
                {\includegraphics[width=1.05\linewidth]{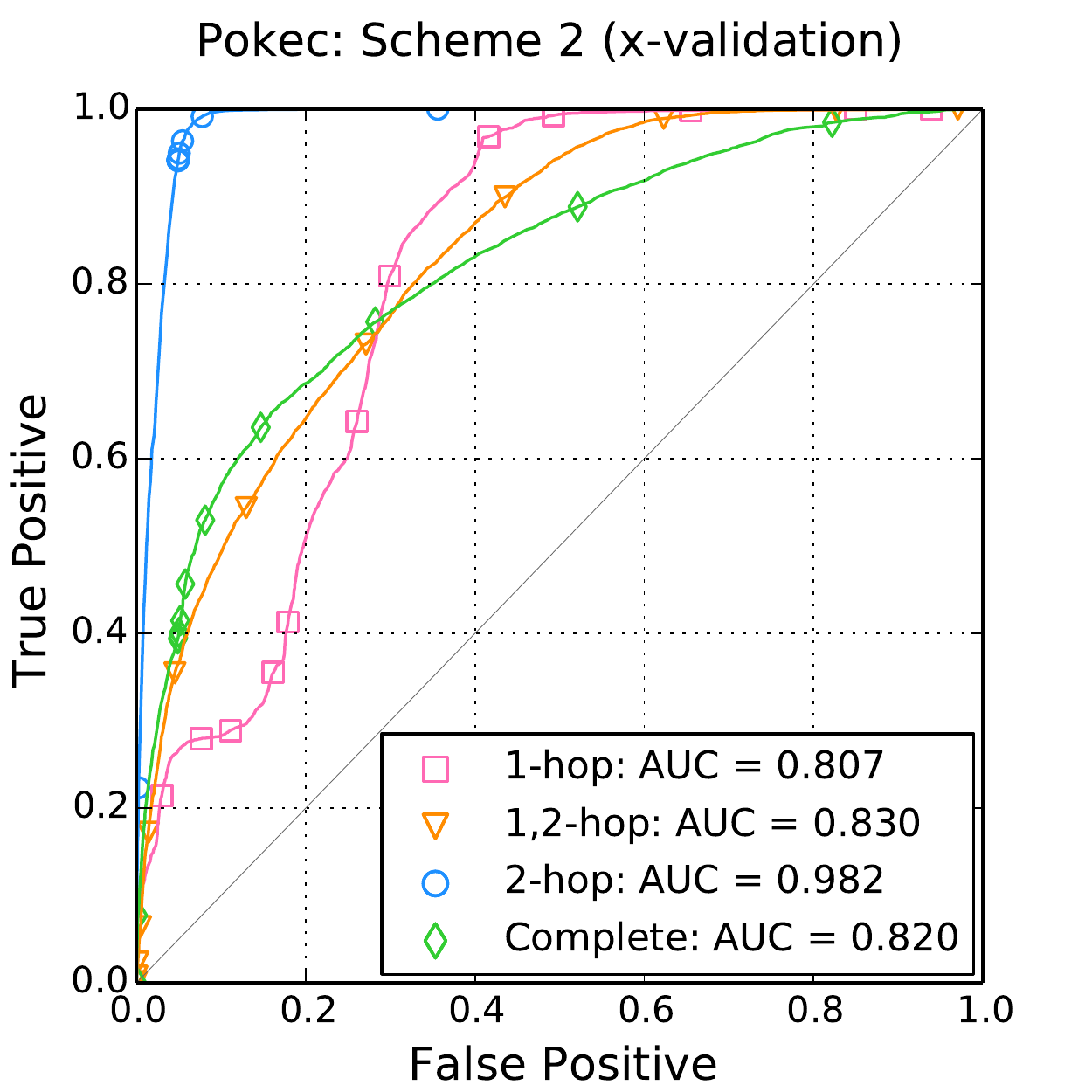}}
            \end{subfigure}%
            \caption{Pokec (x-validation): ROC curves for both schemes}
            \label{fig:roc_pok_x}
        \end{figure}

        \begin{figure}[h!]
            \centering
            \begin{subfigure}{.5\linewidth}
                \centering
                {\includegraphics[width=1.05\linewidth]{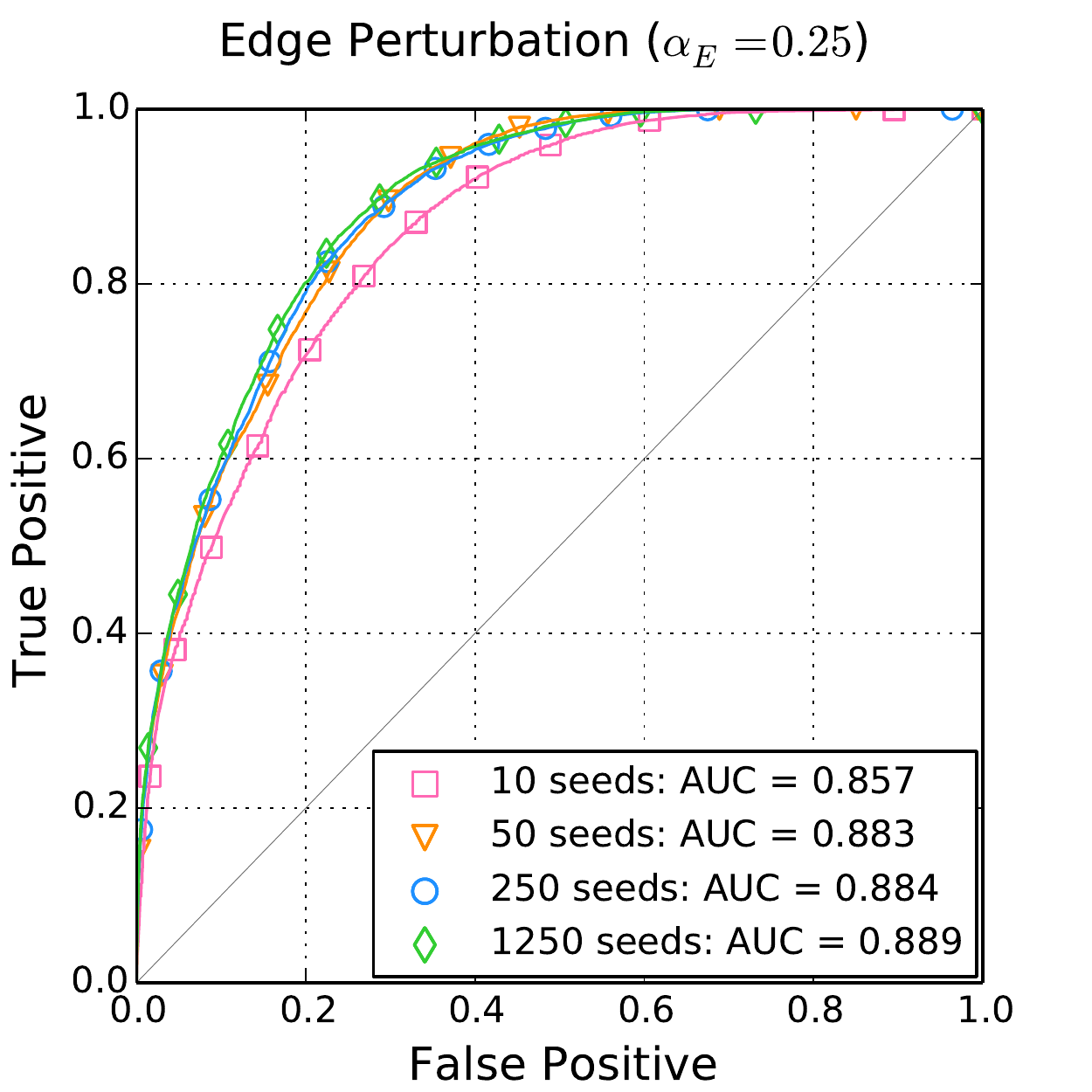}}
            \end{subfigure}%
            \begin{subfigure}{.5\linewidth}
                \centering
                {\includegraphics[width=1.05\linewidth]{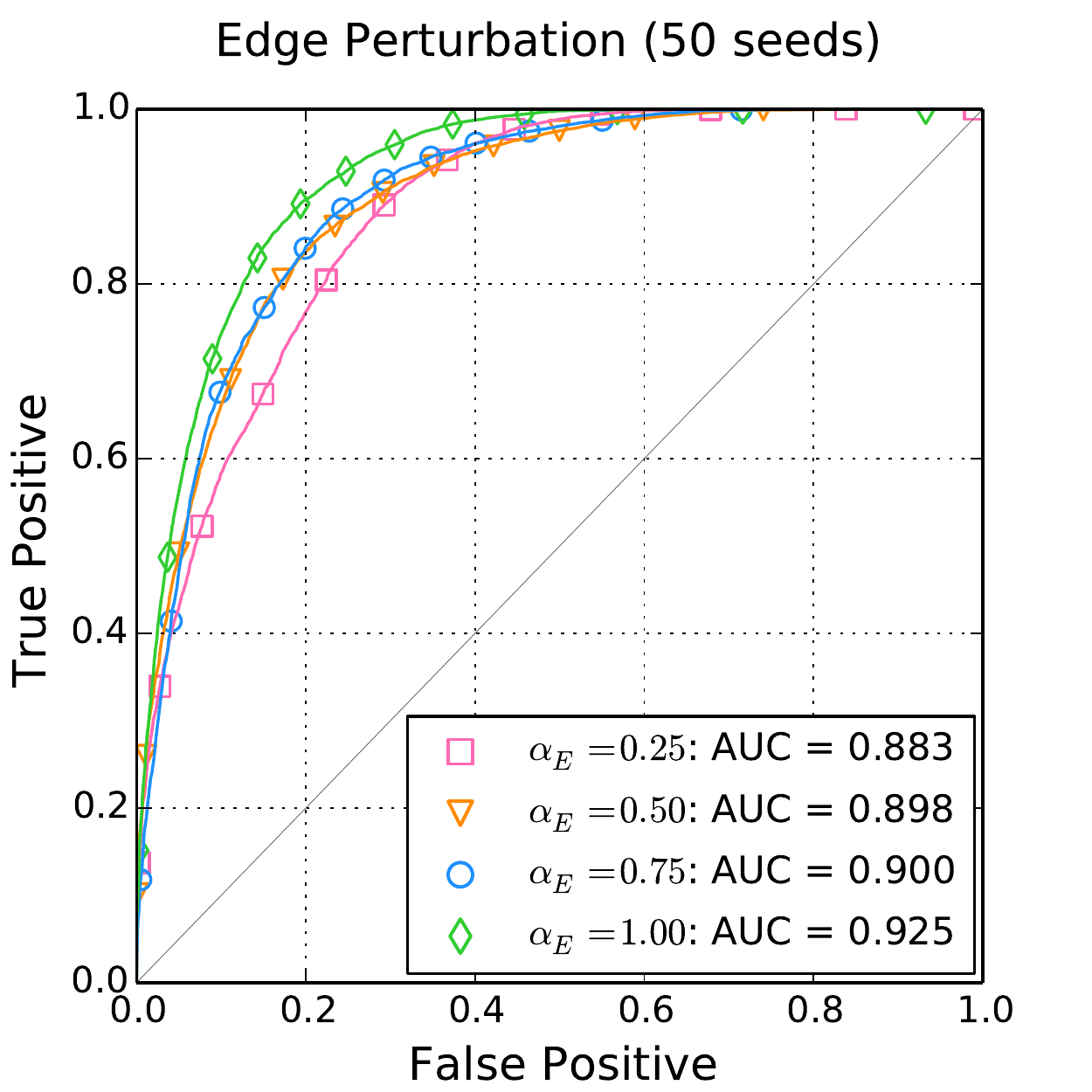}}
            \end{subfigure}%
            \caption{Flickr: ROC curves for edge perturbation}
            \label{fig:roc_flickr}
        \end{figure}

        \begin{figure}[h!]
            \centering
            \begin{subfigure}{.5\linewidth}
                \centering
                {\includegraphics[width=1.05\linewidth]{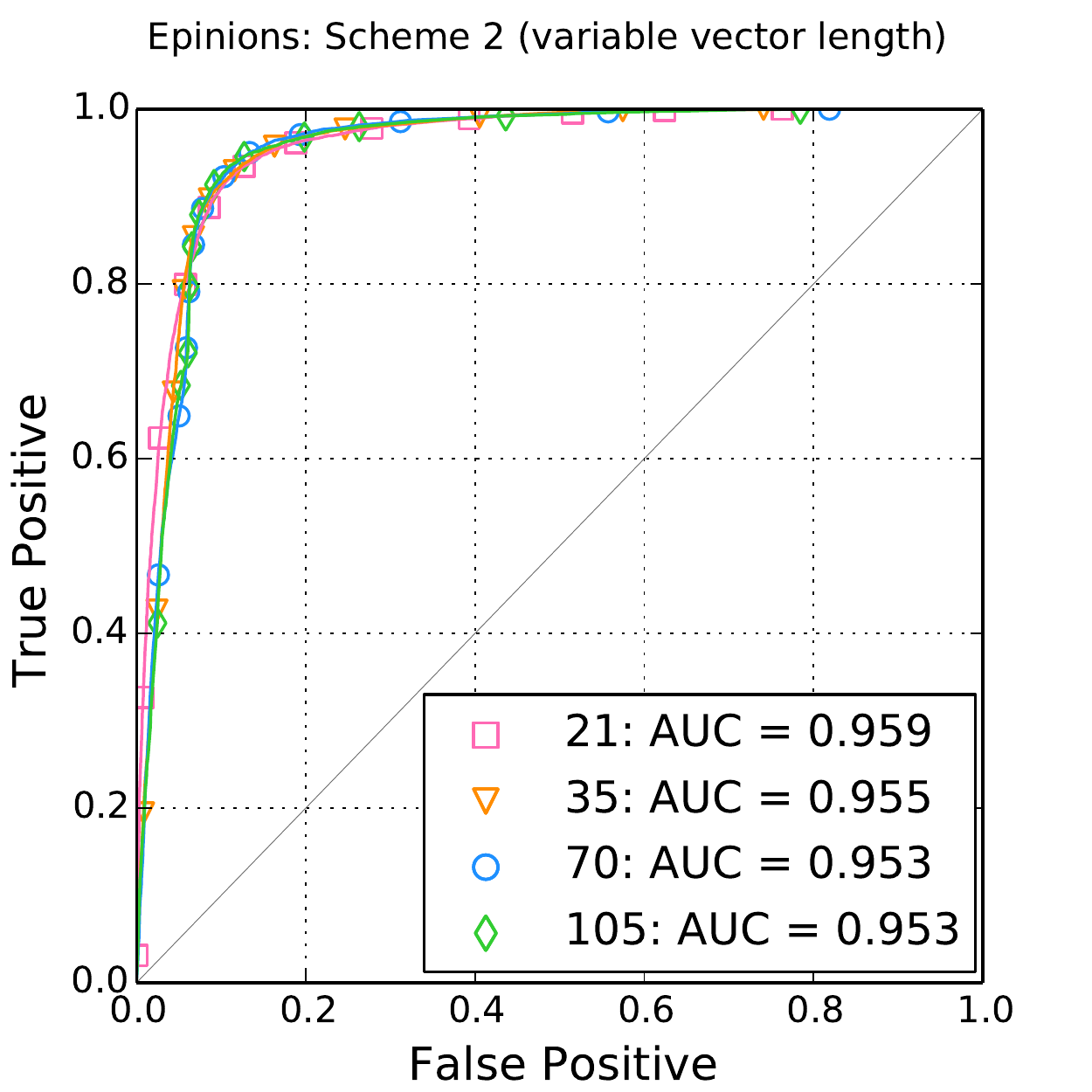}}
            \end{subfigure}%
            \caption{Epinions (Scheme~2, complete): ROC curves for effect of vector length}
            \label{fig:roc_epi_vec_len}
        \end{figure}
\end{document}